\documentclass[manuscript]{aastex}

\usepackage{natbib}
\usepackage{soul}
\newcommand{\micitat}[1]{~\citet{#1}}
\newcommand{\micitap}[1]{~\citep{#1}}
\newcommand{\micitaalt}[1]{~\citealt{#1}}
\newcommand{\kepler}{{\it Kepler}}

\shorttitle{The Planetary System to KIC 11442793}
\shortauthors{Cabrera et al.}

\begin{document}


\title{The Planetary System to KIC 11442793: A Compact Analogue to the Solar System}


\author{
  J.~Cabrera\altaffilmark{1}, 
  Sz.~Csizmadia\altaffilmark{1},
  H.~Lehmann\altaffilmark{2},
  R.~Dvorak\altaffilmark{3},
  D.~Gandolfi\altaffilmark{4,5},
  H.~Rauer\altaffilmark{1,6},
  A.~Erikson\altaffilmark{1},
  C.~Dreyer\altaffilmark{1},
  Ph.~Eigm{\"u}ller\altaffilmark{1} and
  A.~Hatzes\altaffilmark{2}
}
%\affil{Institute of Planetary Research, German Aerospace Center, Rutherfordstrasse 2, 12489 Berlin, Germany}
%\affil{Th{\"u}ringer Landessternwarte, 07778 Tautenburg, Germany}
%\affil{Research and Scientific Support Department, ESTEC/ESA, PO Box 299 2200 AG Noordwijk, The Netherlands}
%\affil{Universit{\"a}tssternwarte Wien, T{\"u}rkenschanzstr. 17, 1180 Wien, Austria}
%\affil{Center for Astronomy and Astrophysics, TU Berlin, Hardenbergstr. 36, 10623 Berlin, Germany}

%%{\bf{Davide now has an affiliation with INAF - ask him what affiliation he needs.}}

\altaffiltext{1}{Institute of Planetary Research, German Aerospace Center, Rutherfordstrasse 2, 12489 Berlin, Germany}
\altaffiltext{2}{Th{\"u}ringer Landessternwarte, 07778 Tautenburg, Germany}
\altaffiltext{3}{Universit{\"a}tssternwarte Wien, T{\"u}rkenschanzstr. 17, 1180 Wien, Austria}
\altaffiltext{4}{Research and Scientific Support Department, ESTEC/ESA, PO Box 299 2200 AG Noordwijk, The Netherlands}
\altaffiltext{5}{INAF - Catania Astrophysical Observatory, Via S.Sofia 78, 95123 Catania, Italy}
\altaffiltext{6}{Center for Astronomy and Astrophysics, TU Berlin, Hardenbergstr. 36, 10623 Berlin, Germany}


% \altaffiltext{1}{Visiting Astronomer, Cerro Tololo Inter-American Observatory.
% CTIO is operated by AURA, Inc.\ under contract to the National Science
% Foundation.}
% \altaffiltext{2}{Society of Fellows, Harvard University.}
% \altaffiltext{3}{present address: Center for Astrophysics,
%     60 Garden Street, Cambridge, MA 02138}
% \altaffiltext{4}{Visiting Programmer, Space Telescope Science Institute}
% \altaffiltext{5}{Patron, Alonso's Bar and Grill}


\begin{abstract}
% Finding planetary systems similar to our own is one of the main goals
% of exoplanet search.  
% It is of particular interest if such systems show planetary transits,
% since multiple transiting planetary systems provide crucial
% information for the understanding of planet formation and
% evolution\micitap{ford2006a}. 
% Mutual dynamical interactions between planets especially require an
% additional effort to understand their origin and to justify their long
% term stability. 
% Unfortunately, such systems are difficult to find because of the low 
% geometrical probability for transiting planets. 
%% {\bf{Known such systems are not numerous because it is difficult to
%% find them owing to the low geometrical probability for transiting planets.}}
We announce the discovery of 
a planetary system with 7 transiting planets around a
\kepler~target, a current record for transiting systems. 
Planets b, c, e and f are reported for the first time in this work.  
Planets d, g and h were previously reported in the
literature\micitap{batalha2013}, although here we revise their
orbital parameters and validate %confirm
their planetary nature.
Planets h and g are gas giants and show strong dynamical interactions.
The orbit of planet g is perturbed in such way that its orbital period
changes by 25.7h between two consecutive transits during the length of
the observations, which is the largest such perturbation found so
far. 
The rest of the planets also show mutual interactions: planets d, e
and f are super-Earths close to a mean motion resonance chain (2:3:4),
and planets b and c, with sizes below 2 Earth radii, are within 0.5\%
of the 4:5 mean motion resonance.  
This complex system presents some similarities to our Solar System,
with small planets in inner orbits and gas giants in outer orbits.
It is, however, more compact. 
The outer planet has an orbital distance around 1 AU, and the relative
position of the gas giants is opposite to that of Jupiter and Saturn,
which is closer to the expected result of planet formation theories.
The dynamical interactions between planets are also much richer.
\end{abstract}





%
%________________________________________________________________
\section{Introduction}
\label{sec:intro}

Finding planetary systems similar to our own is one of the main goals
of exoplanet search.  
It is of particular interest if such systems show planetary transits,
since multiple transiting planetary systems provide crucial
information for the understanding of planet formation and
evolution\micitap{ford2006a}. 
Mutual dynamical interactions between planets especially require an
additional effort to understand their origin and to justify their long
term stability. 
Unfortunately, such systems are difficult to find because of the low 
geometrical probability for transiting planets. 
The satellite \kepler\micitap{borucki2010a} has observed the planetary
system orbiting the star KIC~11442793 almost continuously for more than
4 years.
The \kepler~team has published the parameters of 3 transiting
candidates around this star\micitap{batalha2013} with the
identification numbers KOI 351.01, .02, and .03.  
% We provide in this paper revised parameters for this candidates and an
% indirect confirmation of their planetary nature.
A careful analysis of the light curve with the transit detection
algorithm DST\micitap{cabrera2012} reveals the presence of 4 additional
transiting planets, making this system the most populated among the
transiting ones. %, with one planet more than previous discoveries (i.e. Kepler-11\micitaalt{lissauer2011a}).
 These 4 planets are reported here for the first time (see the 
  results of\micitaalt{ofir2013a,huang2013,tenenbaum2013}\footnote{while this paper was in referee process, Schmitt et al. submitted to AJ a paper with an independent characterization of this system.}).
Considering the magnitude of the star (13.7 magnitude in SDSS r) and 
the characteristics of the transiting candidates, we were not able to
independently confirm the planets by measuring their masses with
radial velocity. 
However, we have performed the following steps to validate %confirm
the planetary nature of the candidates: 
1) medium resolution spectra of the star were taken with the
Coud{\'e}-Echelle spectrograph at the Tautenburg observatory,
characterizing the host star as a solar-like dwarf;
2) the analysis of the \kepler~photometry, including the study of the 
motion of the PSF centroid\micitap{batalha2010b}, which does not
reveal any hint of the presence of contaminating eclipsing binary; 
3) the analysis of the timing of the eclipses reveals that the
planetary candidates are dynamically interacting one with each other;  
and finally 4) a stability analysis of the system with the orbital
dynamics integrator {\em Mercury}\micitap{chambers1999} reveals that,
for the system to be stable, all the planetary candidates must have
planetary masses.  
Therefore, we validate %confirm
in this paper the planetary nature of the 7 candidates. 

%% Section~\ref{sec:star} describes the stellar characterization,  
%% Section~\ref{sec:lightcurve} reports the analysis of the light curve,  
%% Section~\ref{sec:nature} contains the study of the system stability,  
%% and finally Section~\ref{sec:discussion} puts the system into context
%% and discusses the consequences of this discovery. 

%
%________________________________________________________________
\section{Stellar characterization}
\label{sec:star}

In order to characterize the host star, five spectra were taken on
June 6 and 7, 2013, with the Coud{\'e}-Echelle spectrograph attached
to the 2-m telescope at the Th{\"u}ringer Landessternwarte Tautenburg.
The wavelength coverage was 472-736 nm and a 2 arcsec slit provided a
spectral resolving power of $32\,000$. 
The exposure time for each spectrum was 40 minutes.

The spectra were reduced using standard ESO-MIDAS packages. 
The reduction steps included filtering of cosmic rays, background and
straylight subtraction, flat fielding using a halogen lamp, optimum
extraction of diffraction orders, and wavelength calibration using a
ThAr lamp.  
Due to the low signal-to-noise ratio (SNR) of a single spectrum it was
difficult to define the local continuum.  
Because no radial velocity shifts between the single spectra could be
found, we repeated the reduction using the co-added raw spectra.  
%% \footnote{{\bf{Considering the one day time-interval between these spectra,
%% and the expected low RV-signal (cf. Section 5), this fact does not contradict
%% the planetary nature of the transiting objects.}}}
The continuum of the resulting mean spectrum was then well enough
defined for a proper normalization.  
The SNR of the mean spectrum, measured from some almost line-free
parts of the continuum, was about 19. 

We used the spectral synthesis method, which compared the observed
spectrum with synthetic spectra computed on a grid in atmospheric
parameters. 
The synthetic spectra were computed with the SynthV
program\micitap{tsymbal1996}, based on a library of atmosphere models
calculated with the LLmodels code\micitap{shulyak2004}. 
The error estimation was done from $\chi^2$ statistics taking all
interdependencies between the different parameters into
account\micitap{lehman2011}. 
%% (see\micitaalt{lehmann2011} for a description of the method). 

The step widths of the grid were 100K in T$_{\mathrm{eff}}$, 0.1 dex in
$\log g$, 0.1 dex in [M/H], 0.5 km\,s$^{-1}$ in microturbulent velocity, and 1
km\,s$^{-1}$ in $v \sin i$; where [M/H] means scaled solar abundances. 
For the determination of $v \sin i$ we used the metal lines-rich
wavelengths region 491-567 nm. 
For all other parameters, the  wavelengths range utilized was 
\mbox{472-567\,nm} which also includes H$_{\beta}$. 

Table~\ref{table:star} lists the results obtained from the full grid
in all parameters. 
% The large errors mainly come, besides from the low SN of the observed
% spectrum, from the large ambiguities between the different parameters. 
The large uncertainties mainly originate from the large ambiguities
between the different parameters and from the low SNR of the observed
spectrum. 
% Assuming that the star is a typical main sequence star of early G-type
% (see below), we can assume $\log g$ of 4.4. 
% Fixing this parameter in the analysis to that value, we get a better
% constraint on $T_{\mathrm{eff}}$ and a slightly higher
% metallicity (last columnn of Table~\ref{table:star}). 
We use a compilation of empirical values of stellar parameters
from\micitap{gray2005}. 
Comparing our results from the full grid search with the literature
data, we see that we can exclude luminosity class III stars because of
the values of $\log g$ and v$\sin i$. 
The $T_{\mathrm{eff}}$ derived from spectral analysis lies, within the
uncertainties, between $5\,600$ and $6\,250$K which is consistent with
dwarfs of spectral type G6 to F6.
Based on the measured $\log g$, we cannot determine if the star is
slightly evolved. 
Assuming that the star is a typical main sequence star of early
G-type, we can adopt a $\log g$ of 4.4, which lies within the
measurement error, obtaining a better constraint on on
$T_{\mathrm{eff}}$ and a slightly higher value for the metallicity
(last column of Table~\ref{table:star}).
Under this assumption, we obtain $T_{\mathrm{eff}}$ between $5\,910$
and $6\,340$ K, corresponding to spectral types between G1 and F6. 
The corresponding ranges in mass and radius are relatively small, 
between 1.1 and 1.3 M$_{\mathrm{sun}}$ and 1.1 and 1.3
R$_{\mathrm{sun}}$. 
 
\subsection{Reddening and distance}
\label{subsec:extinction}

\begin{figure}
  \centering
  \includegraphics[%
    width=0.9\linewidth,%
    height=0.5\textheight,%
    keepaspectratio]{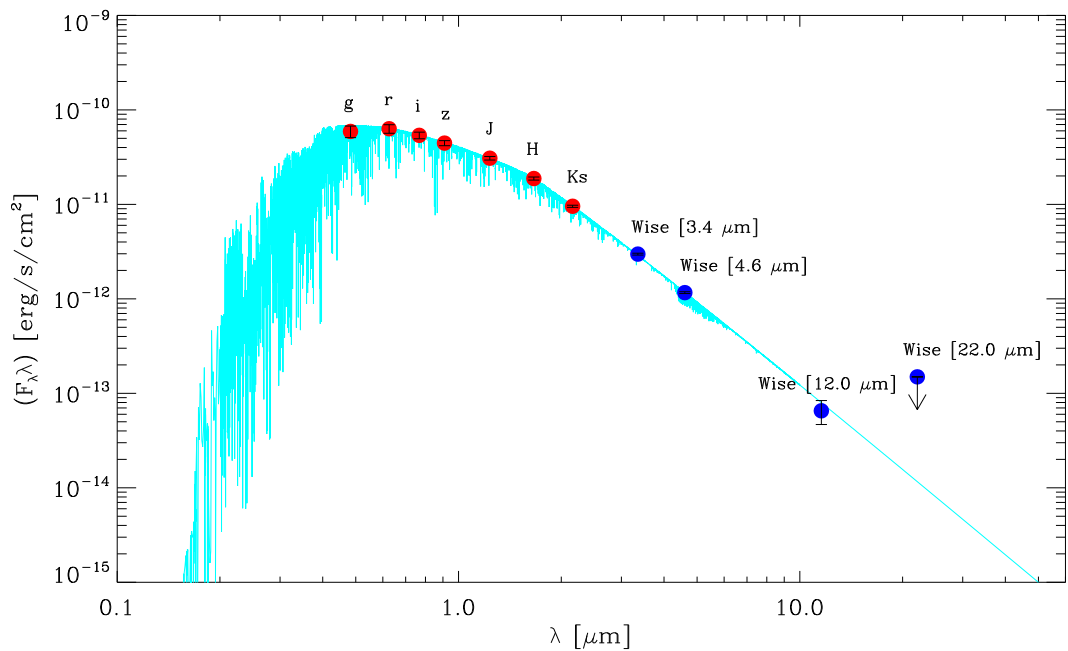}
  \caption{Dereddened spectral energy distribution of
    KIC\,11442793. 
    The optical SDSS-$g$,-$r$,-$i$,-$z$ photometry is from the
    \kepler~Input Catalogue.
    Infrared $J$,$H$,$Ks$ and $W1$, $W2$, $W3$, $W4$ data are taken
    from the 2MASS\micitap{cutri2003} and WISE\micitap{wright2010wise}
    database, respectively. 
    The \emph{NextGen} model spectrum by\micitat{hauschildt1999a} with
    the same photospheric parameters as KIC\,11442793 and scaled to
    the stellar radius and distance is overplotted with a light-blue
    line.} 
\label{figure:sed}
\end{figure}

We determined the interstellar extinction $A_\mathrm{v}$ and distance
$d$ to KIC\,11442793 by applying the method described 
in\micitat{gandolfi2008}. 
This technique is based on the simultaneous fit of the observed
stellar colours with theoretical magnitudes obtained from the
\emph{NextGen} model spectrum\micitap{hauschildt1999a} with the   
same photospheric parameters as the target star. 
For KIC\,11442793 we used SDSS, 2MASS, and WISE
photometry (see Table~\ref{table:magnitudes} and Fig.~\ref{figure:sed}). 
We excluded the $W_3$ and $W_4$ WISE magnitudes, as the former has a
SNR of 3.5 and the latter is only an upper limit. 
Assuming a normal extinction ($R_\mathrm{v}=3.1$) and a black body
emission at the stellar effective temperature and radius, we found
that the star reddening amounts to $A_\mathrm{v}=0.15\pm0.10$\,mag and
that the distance to KIC\,11442793 is $d=780\pm100$\,pc.  

\begin{table}
  \caption[]{Derived atmospheric parameters for the star.}
  \label{table:star}
  \centering
  \begin{tabular}{lcc}
          & full grid & $\log g$ fixed \\
    \hline\hline
    T$_{\mathrm{eff}}$ (K)             & $5\,930 \pm 320$  & $6\,080^{+260}_{-170}$ \\
    $\log g$ (cgs)                   &  $4.0   \pm 0.5$  &  $4.4$ (fixed)       \\
    v$_{\mathrm{mic}}$ (km\,s$^{-1}$)  &  $1.2   \pm 0.6$  &  $1.2  \pm 0.6$      \\
    $[$M/H$]$ (dex)                  & $-0.17  \pm 0.21$ & $-0.12 \pm 0.18$     \\
    v $\sin i$ (km\,s$^{-1}$)        &  $4.6   \pm 2.1$  &  $4.6  \pm 2.1$      \\
    \hline
  \end{tabular}
\end{table}

\begin{table}
  \caption{\kepler, GSC2.3, USNO-A2, and 2MASS identifiers of the target star. 
           Equatorial coordinates and optical SDSS-$g$,-$r$,-$i$,-$z$ photometry are from
           the \kepler~Input Catalogue. Infrared $J$,$H$,$Ks$ and $W1$,$W2$,$W3$,$W4$ data are
           taken from the 2MASS\micitap{cutri2003} and
           WISE\micitap{wright2010wise} database, respectively.} 
  \label{table:magnitudes}      
  \begin{center}
  \begin{tabular}{lll}       
  \multicolumn{1}{l}{\emph{Main identifiers}}     \\
  \hline
  \hline
  \noalign{\smallskip}                
  \kepler~IDs     & KIC~11442793 - KOI~351 - Kepler-90 \\
%%  {\bf Kepler Object of Interest} & {\bf KOI 351} \\
  GSC2.3~ID       & N2EM001018        \\
  USNO-A2~ID      & 1350-10067455     \\
  2MASS~ID        & 18574403+4918185  \\
  \noalign{\smallskip}                
  \hline
  \noalign{\medskip}
  \noalign{\smallskip}                
  \multicolumn{2}{l}{\emph{Equatorial coordinates}}     \\
  \hline                                  
  \hline                                  
  \noalign{\smallskip}                
%  RA \,(J2000)      & $18^h\,57^m\,44^s.038$         \\
%  Dec (J2000)       & $+49^\mathrm{o} 18' 18''.58$  \\
  RA \,(J2000) $18^h\,57^m\,44^s.038$ & Dec (J2000)  $+49^\mathrm{o} 18' 18''.58$  \\
  \noalign{\smallskip}                
  \hline
  \noalign{\medskip}
  \noalign{\smallskip}                
  \multicolumn{3}{l}{\emph{Magnitudes}} \\
  \hline
  \hline
  \noalign{\smallskip}                
  \centering
  Filter \,\,($\lambda_{\mathrm eff}$)& Mag     & Uncertainty  \\
  \noalign{\smallskip}                
  \hline
  \noalign{\smallskip}                
  $g$ \,\,~\,(~0.48\,$\mu m$) & 14.139       & 0.030 \\
  $r$ \,\,~\,(~0.63\,$\mu m$) & 13.741       & 0.030 \\
  $i$ \,\,~\,(~0.77\,$\mu m$) & 13.660       & 0.030 \\
  $z$ \,\,~\,(~0.91\,$\mu m$) & 13.634       & 0.030 \\
  $J$ \,\,~\,(~1.24\,$\mu m$) & 12.790       & 0.029 \\
  $H$ \,\,\,(~1.66\,$\mu m$)  & 12.531       & 0.033 \\
  $Ks$  \,(~2.16\,$\mu m$)    & 12.482       & 0.024 \\
  $W_1$ \,(~3.35\,$\mu m$)    & 12.429       & 0.024 \\
  $W_2$ \,(~4.60\,$\mu m$)    & 12.462       & 0.024 \\
  $W_3$ \,(11.56\,$\mu m$)    & 12.750       & 0.308 \\
  $W_4$ \,(22.09\,$\mu m$)    &$~~9.702^{a}$ &   ~~~-\\
  \noalign{\smallskip}                
  \hline
  \end{tabular}
  \end{center}
  {$^{a}${Upper limit}}
\end{table}

%
%________________________________________________________________
\section{Light curve analysis}
\label{sec:lightcurve}

\begin{figure}
  \centering
  \includegraphics[%
      width=0.9\linewidth,%
      height=0.5\textheight,%
      keepaspectratio]{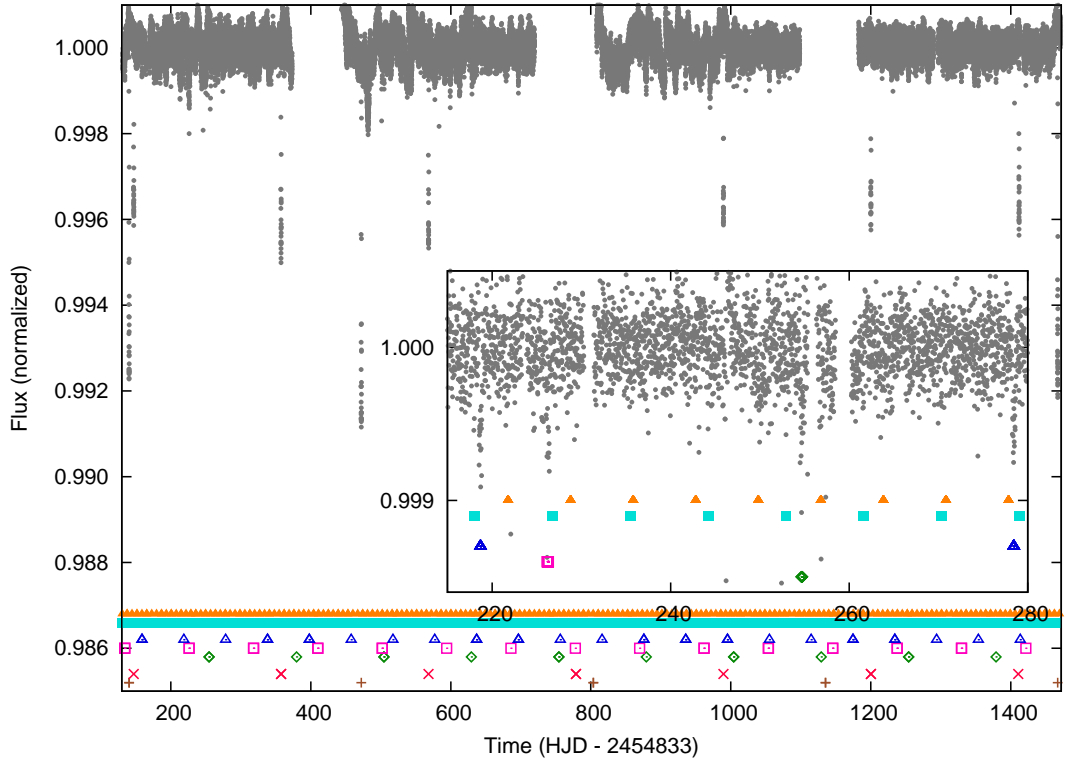}
  \caption{Public raw light curve of KIC~11442793. The seven sets of
    periodic transits are indicated with symbols of different colors:
    planet h with red plus-signs, planet g with red crosses, planet f
    with green diamonds, planet e with magenta squares, planet d with
    blue triangles, planet c with turquoise filled squares, and planet
    b with orange filled triangles. In the enlarged region the stellar 
    variability has been subtracted to show a subset of the shallower 
    transits.}
  \label{figure:rawlc}
\end{figure}

\kepler~observations of KIC~11442793 extend for $1\,340$ days with a
duty cycle of 82\%. 
% faltan: 
% 70d entre  373 y  443
% 88d entre  719 y  808
% 84d entre 1098 y 1182
% (70.+88.+84.)/1340. = 0.18
%% There are several interruptions due to the failure of one of the CCDs
%% of the Kepler instrument {\bf (reference?).}
The light curve, shown in Figure~\ref{figure:rawlc}, reveals that the
host star is not particularly active. 
It barely shows hints of some variations compatible with the evolution
of stellar spots on its surface, with an amplitude of 0.1\%  

We have applied a detrending algorithm to treat the stellar activity
optimized for the CoRoT mission\micitap{baglin2006}, but adapted to
the treatment of \kepler~data\micitap{cabrera2012}.
Then we have applied the transit detection algorithm
DST\micitap{cabrera2012} to search for the periodic signature of 
transiting planets.

We confirm the detection of the candidates KOI~351.01, .02, and .03,
previously announced\micitap{batalha2013}, and we assign them the
identifications KIC~11442793~h, g, and d. 
We present the discovery of four additional candidates, b, c, e, and
f, reported for the first time here.   
The ephemerides of these objects are given in Table~\ref{table:planets}. 
The orbital ephemerides have been calculated as follows:
the transit detection algorithm DST provides preliminary values of the
period, epoch, depth and duration of the transiting candidates.
With this information, we first fit separately the transits of every
candidate. 
Then we make a weighted linear fit to the epochs of the individual
transits, the slope of the fit is the period and the intercept the
epoch.% shown in Table~\ref{table:planets}.
The residuals between the linear fit and the actual position of the
transits (observed minus calculated, O-C) are usually referred as
transit timing variations (TTVs), which are discussed later on
Section~\ref{sec:ttv}.

%
%________________________________________________________________
\section{Planetary parameter modelling}
\label{sec:szilard}

Several planets in this system show significant transit timing
  variations (TTVs), described in Section~\ref{sec:ttv}, which need to
  be removed before proceeding with the modeling of the planetary
  parameters.
  We use an iterative method to correct for this effect, similar to
  the one described by\micitat{alapini2009}, but accounting for the
  TTVs. 
  We take a geometrical model of the transit based on the preliminary
  value of the planetary parameters obtained by the detection method.
  We use a genetic algorithm\micitap{geem2001} to fit the value of the
  epoch, fixing the other transit parameters.
  For every trial value of the epoch, we correct for stellar activity
  in a region covering ten times the transit duration with a second
  order Legendre polynomial (first order for planets b and c).
  The polynomial is interpolated in the expected region of the
  transit, to preserve the transit shape.
  We then fold the light curve with the obtained values of the
  individual epochs.
  This method does not converge in the case of planets b and c because
  of the low SNR of their transit signal.
  Therefore, for planets b and c we fix the period, we do not fit for
  the epochs, but we do apply the stellar activity correction for each
  individual transit described above.

%% The publicly available Kepler light curves, processed by the Kepler
%% data pipeline, were postprocessed in the same way as
%% in\micitat{csizmadia2011} to remove the stellar variability as well as
%% remaining instrumental effects. 
%% This consists of the following steps. 
%% First, we selected the segments of the light curves of every transit,
%% each segments starting 2D before the mid-transit and ended 2D after
%% the mid-transit, where D is the transit duration. 
%% Each segment was separately fitted by a parabola using only the
%% out-of-transit points, and finally the whole segment was divided by
%% the parabola. 
%% We used only the short-cadence data.
%% Then we selected those transits which exhibited {\st{negligible}} 
%% {\bf {small}} TTV-values {\bf {and we shifted the transits to the same phases.
%% - IF I UNDERSTAND YOU WELL}}
%% {\st{and}} {\bf{Then}} we folded the light curves with the average periods. 
%% For candidates b, c, d, e and f we binned the light curves (we formed
%% {\bf{2000}} {\st{500}} binned points in the $\pm$2D vicinity of the mid-transit), while
%% for candidates g and h we used the original short cadence photometric
%% points.  
A detailed description of the modeling of the planetary parameters
applied here can be found in\micitat{csizmadia2011}.
We used the publicly available short cadence \kepler~light curves.
For candidates b, c, d, e and f we binned the light curves (we formed
2000 binned points in the $\pm 2D$ vicinity of the mid-transit, $D$
being the transit duration), while for candidates g and h we used the
original short cadence photometric points.  
We used the\micitat{mandel2002} transit model. 
This model gives the light loss of the star due to the transit of an 
object as a function of their size ratio (k), of their mutual
sky-projected distance (denoted by $\delta$), and of the limb
darkening coefficients of the transited star 
($ld_1 = u_1 + u_2,$ and $ld_2 = u_1 - u_2$). 
%% {\st{During the light curve modeling procedure, the limb darkening
%% coefficients were fixed at values based on %\micitat{claret2013}
%% quasi-spherical limb darkening calculations for the Kepler
%% pass-bands. Since it has tabulated values, they were interpolated to the stellar
%% effective temperature and $\log g$ (assuming solar metalicity).}}

Following\micitat{csizmadia2013} we determined the limb darkening
coefficients from the light curve instead of using theoretical
predictions.  
This fit was first applied to planet h, which has the largest transit
depth i.e. the highest signal-to-noise ratio. 
Having obtained the values of the limb darkening coefficients, we set
the limb darkening coefficients at the value obtained from the fit of
planet g's transit light curve, but we allowed them to vary within the
uncertainties of the determined values.

Since we do not have any radial velocity measurements, nor
occultations, nor phase-curves of any of these seven planets, we had
no a priori information about eccentricities and arguments of periastron.
Therefore we could not calculate the sky-projected distance of the
stellar and planetary centers in the usual way
(e.g.\micitaalt{gimenez2006}).
% \begin{equation}
% \delta = \frac{a(1-e^2)}{1+e \cos \nu} \sqrt{ 1 - \sin^2 i \sin^2 (\nu + \omega)}
% \end{equation}
% where $a$ is the semi-major axis, $e$ is the eccentricity, $\nu$ is
% the true anomaly (also a function of eccentricity, orbital period and
% periastron passage time), $i$ is the inclination and $\omega$ is the
% argument of periastron. 
Instead, we fitted the duration of the transit, the epoch ($t_0$), the
period ($P$), the impact parameter ($b$), the planet-to-stellar 
radii ratio ($k=R_p/R_s$). 
Then the sky-projected mutual distance of the star and the planet were
calculated with the formula 
\begin{equation}
\delta \approxeq \sqrt{ b^2 + \left[(1+k)^2 - b^2\right]\left(\frac{t-t_0}{P}\right)^2}
\end{equation}
where $t$ is the time. 
We checked the validity of this latter formula via numerical
experiments and we found that in our cases it yields a very good
agreement with the theoretical value in the vicinities of transits. 
No mutual transit event was modeled.
For the optimization, a genetic algorithm process described
in\micitat{csizmadia2011} was used, and the results were refined by a
Simulated Annealing algorithm which was also used for the error
estimation. 
The reported uncertainties in Table~\ref{table:planets} are 1$\sigma$
uncertainties. 

We report the modeled values of $k$ and $b$ in
Table~\ref{table:planets} with their respective uncertainties for each
of the seven candidates in the system. 
Once $k$, $b$, $D$, $P$ became known from the modeling procedure, the 
value of the scaled semi-major ($a/R_s$) for circular orbits can then
be calculated as 
\begin{equation}
\frac{a}{R_s} = \frac{1}{\pi}\frac{P}{D}\sqrt{(1+k)^2-b^2}
\end{equation}

\begin{figure}%[t]
  \centering
  \begin{minipage}[t]{0.48\textwidth}
    \begin{center}
      \includegraphics[%
        width=0.8\linewidth,%
        height=0.4\textheight,%
        keepaspectratio]{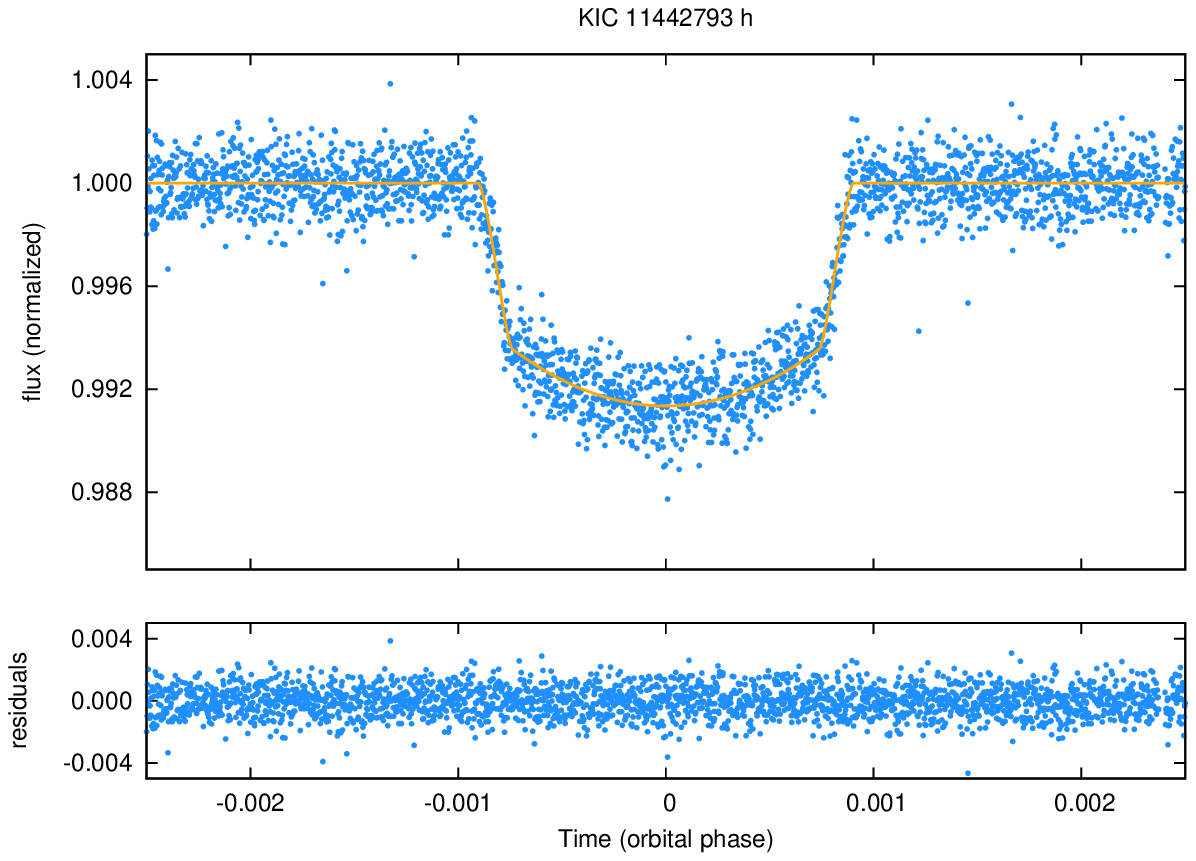}
      \includegraphics[%
        width=0.8\linewidth,%
        height=0.4\textheight,%
        keepaspectratio]{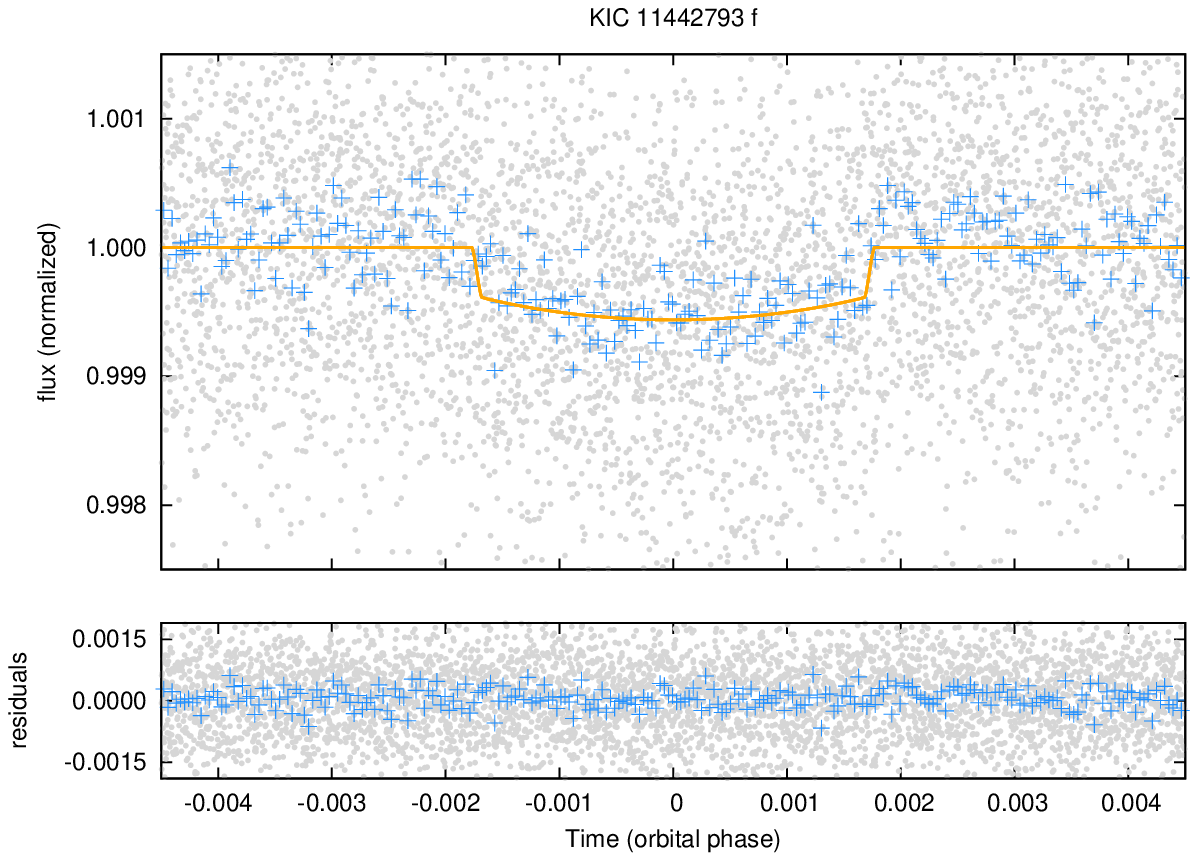}
      \includegraphics[%
        width=0.8\linewidth,%
        height=0.4\textheight,%
        keepaspectratio]{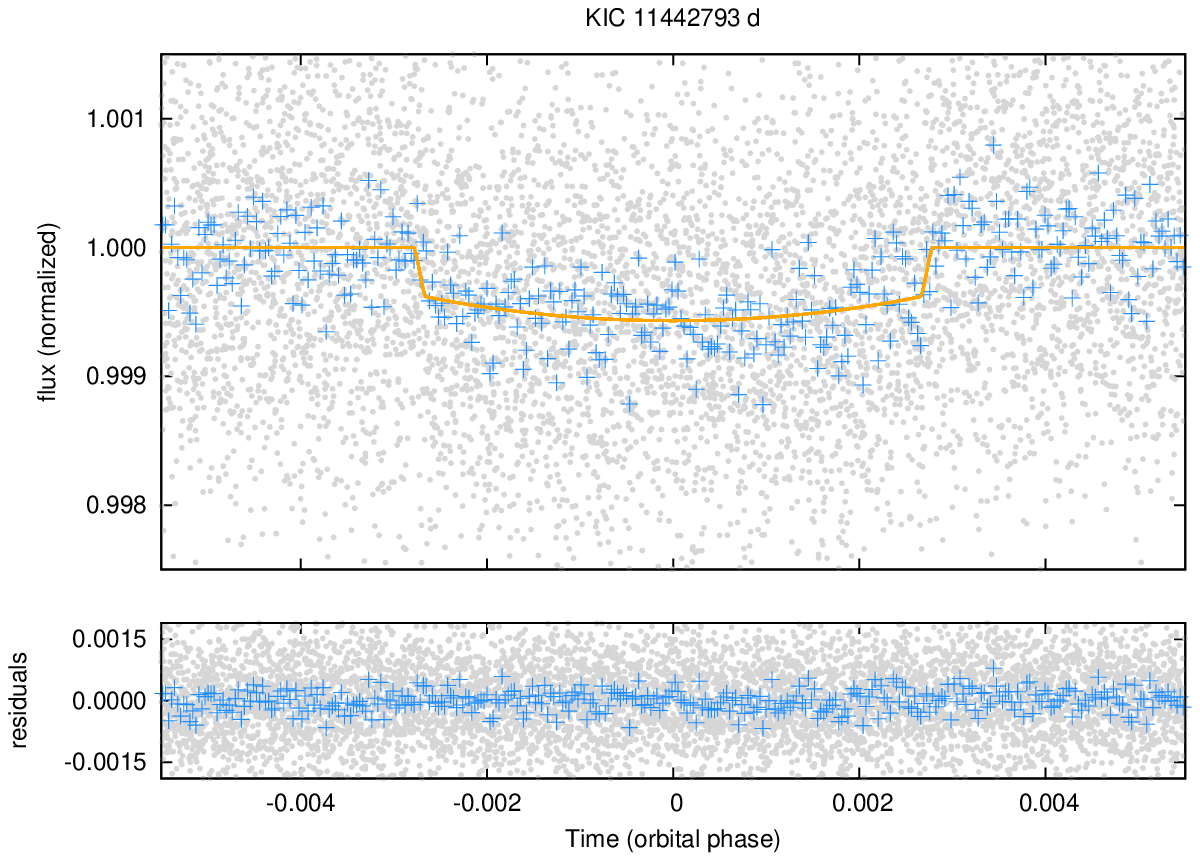}
      \includegraphics[%
        width=0.8\linewidth,%
        height=0.4\textheight,%
        keepaspectratio]{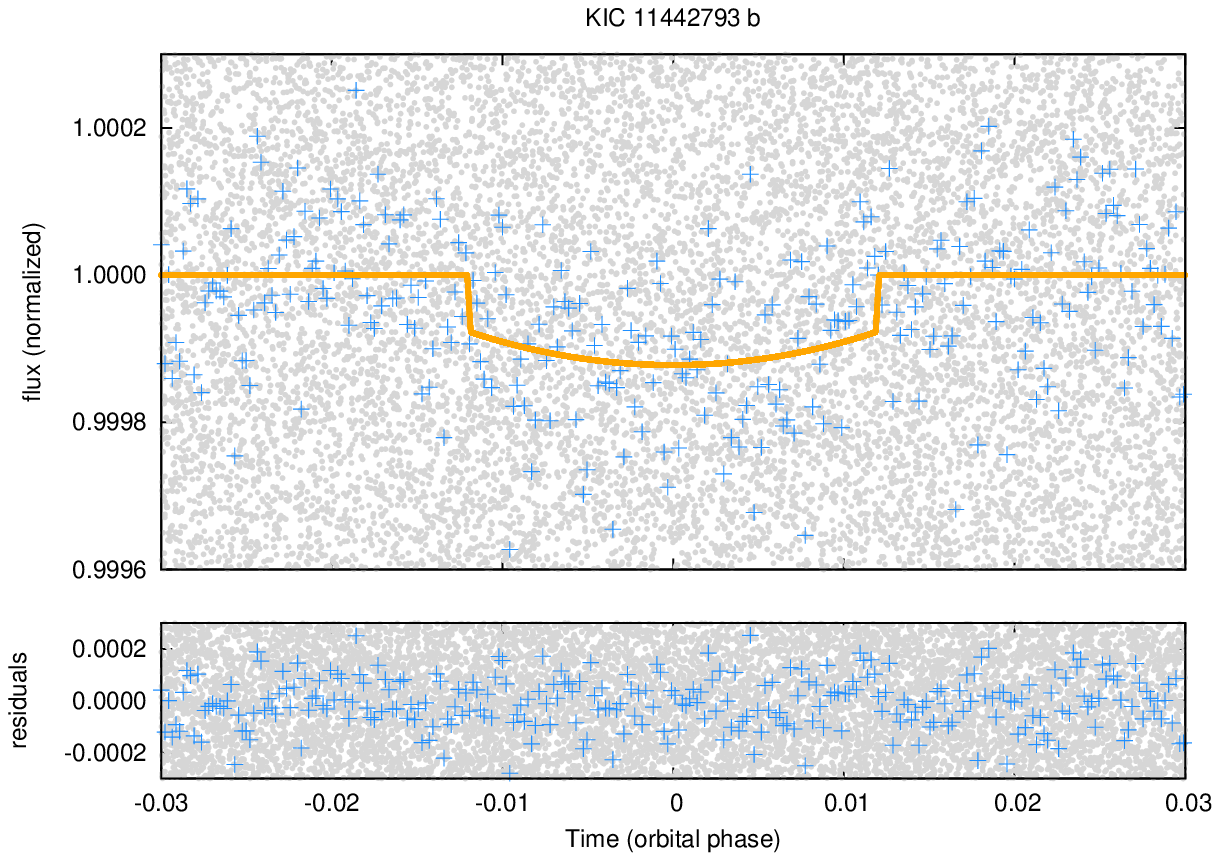}
    \end{center}
  \end{minipage}
  \begin{minipage}[t]{0.48\textwidth}
    \begin{center}
      \includegraphics[%
        width=0.8\linewidth,%
        height=0.4\textheight,%
        keepaspectratio]{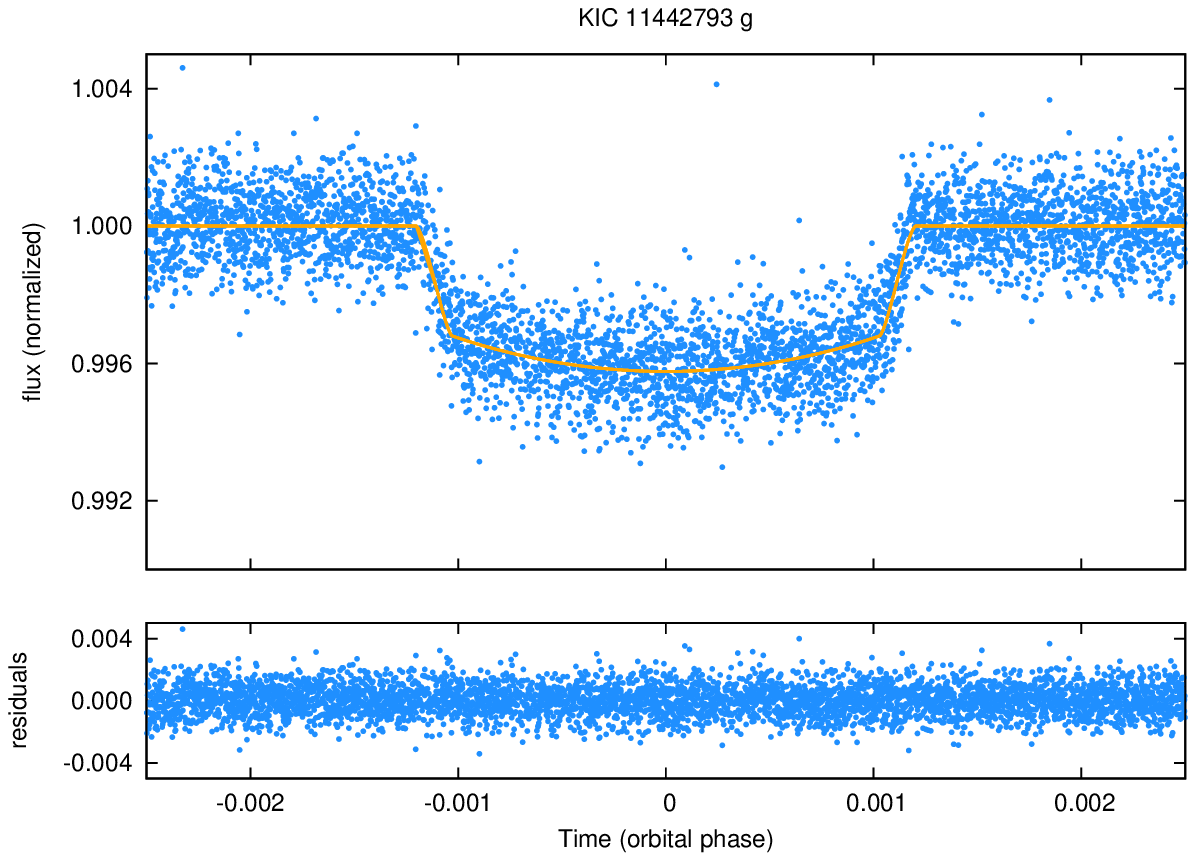}
      \includegraphics[%
        width=0.8\linewidth,%
        height=0.4\textheight,%
        keepaspectratio]{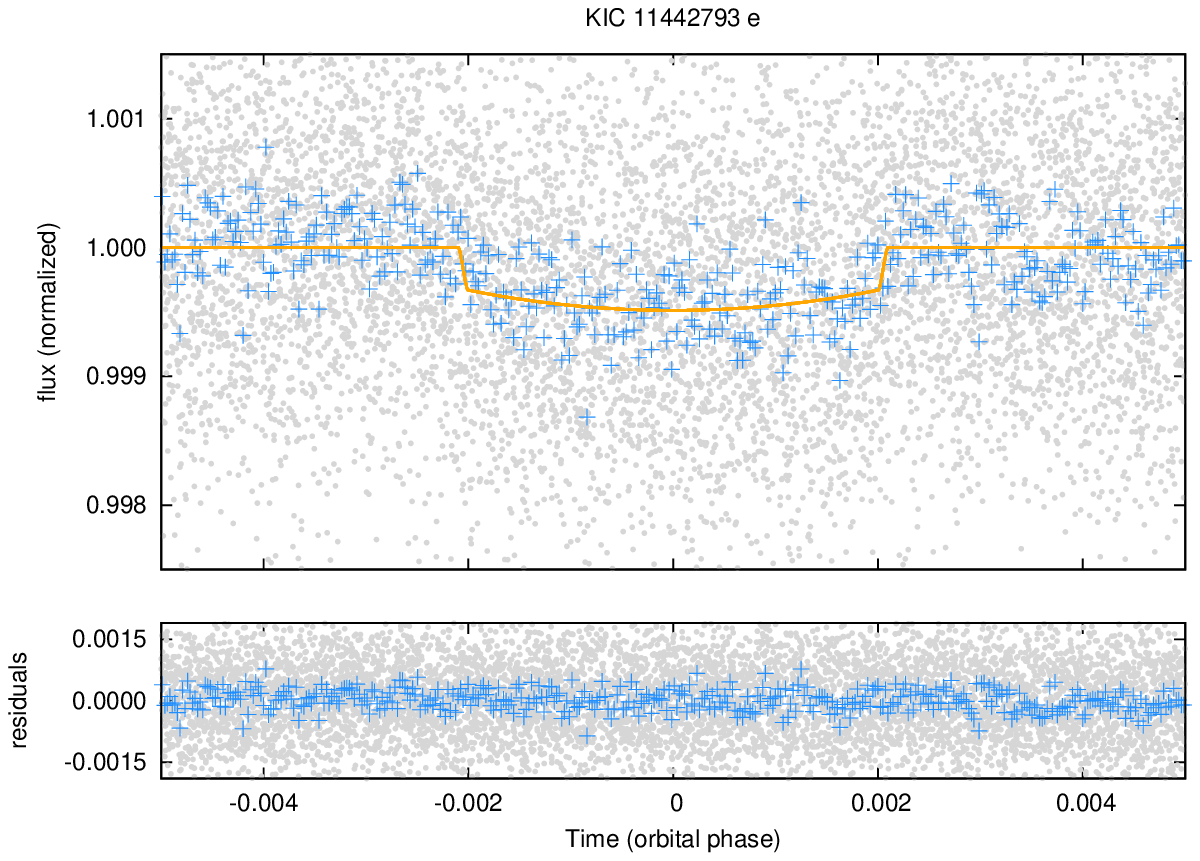}
      \includegraphics[%
        width=0.8\linewidth,%
        height=0.4\textheight,%
        keepaspectratio]{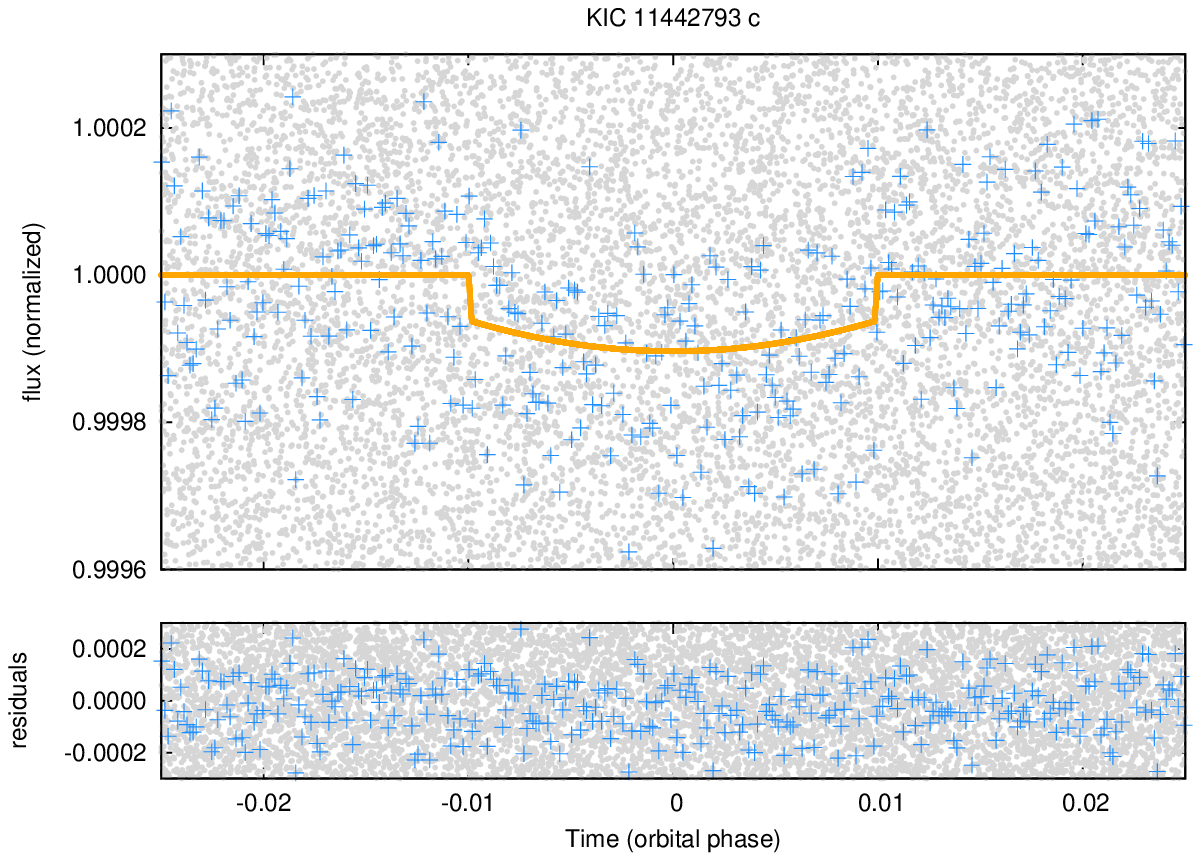}
    \end{center}
  \end{minipage}
  \caption{Filtered light curve of KIC 11442793 folded at the period
    of the different planets. For planets b and c the light curve has
    been binned, to help the eye. 
    The orange solid line shows the light curve fit (Table~\ref{table:planets}). 
    The lower panels show respectively the residuals of the light curve fit.
  }
  \label{figure:fit}
\end{figure}

We then calculated the scaled semi-major axes for every planet in the 
system assuming circular orbits (see Table~\ref{table:planets}). 
Re-writing Kepler's third law, we obtained for the stellar density
parameter (neglecting the mass of the planet):
\begin{equation}
\frac{M^{1/3}}{R_s} = \left( \frac{3\pi}{G P^2} \right)^{1/3} \frac{a}{R_S}
\end{equation}
or equivallentily
\begin{equation}
\frac{M^{1/3}}{R_s} = \left( \frac{3}{G P D} \right)^{1/3} \left[ (1+k)^2-b^2 \right]^{(3/2)} \left( \frac{1-e^2}{1+e^2-2 e \sin \omega} \right)^{3/2}
\end{equation}
We also report the density parameter derived from every candidate in
Table~\ref{table:planets}.
% {\bf Figures~\ref{figure:fith} to~\ref{figure:fitb} show in graphical
 Figure~\ref{figure:fit} shows in graphical
  form the modelling of the photometric light curves and the model
  residuals for each planet.

\begin{table}\footnotesize
  \caption[]{Planetary parameters. Values calculated for $R_s = 1.2\pm0.1 R_\odot$;
    $R_\odot = 696\,342$km and $R_{E} = 6\,378$ km.}
%% {\bf THIS TABLE WAS REVISED BY SZIALRD - 11
%%   Sep 2013 Semi-major axis in AU was calculated from the modelled $a/R_s$ value
%%   and from the assumed $R_s = 1.2\pm0.1 R_\odot$}. $R_\odot = 696 342$km and
%%   $R_{E} = 6 378$ km was used to scale the radius-ratios.}
  \label{table:planets}
  \centering
  \renewcommand{\arraystretch}{0.5}
  \begin{tabular}{*{2}{lc}}
    \hline\hline
    KIC~11442793~h (KOI~351.01) &                          &    KIC~11442793~d (KOI~351.03) & \\
    period (days)         & 331.600\,59  $\pm$ 0.000\,37   &    period (days)         &  59.736\,67  $\pm$ 0.000\,38  \\
    epoch (HJD - 2454833) & 140.496\,31  $\pm$ 0.000\,82   &    epoch (HJD - 2454833) & 158.965\,6   $\pm$ 0.004\,2   \\
    duration (h)          &      14.737  $\pm$ 0.046       &    duration (h)          &        8.40  $\pm$ 0.19      \\
    $a/R_s$               &       180.7  $\pm$ 4.7         &    $a/R_s$               &        56.1  $\pm$ 4.8        \\
    $a$ (AU)              &        1.01  $\pm$ 0.11        &    $a$ (AU)              &        0.32  $\pm$ 0.05       \\
    $R_p/R_s$             &      0.0866  $\pm$ 0.0007      &    $R_p/R_s$             &      0.0219  $\pm$ 0.0005      \\
    $R_p$ ($R_E$)         &        11.3  $\pm$ 1.0         &    $R_p$ ($R_E$)         &        2.87  $\pm$ 0.30        \\
    $b$                   &        0.36  $\pm$ 0.07        &    $b$                   &        0.28  $\pm$ 0.25       \\
    $i$ (deg)            &         89.6 $\pm$ 1.3          &    $i$ (deg)             &        89.71 $\pm$ 0.29       \\
    $M^{1/3}/R_s$         &        0.90  $\pm$ 0.13         &    $M^{1/3}/R_s$         &        0.88  $\pm$ 0.15       \\
    $ld_1$               &        0.348 $\pm$ 0.056        &    $ld_1$               &        0.371 $\pm$ 0.087       \\
    $ld_2$               &         1.03 $\pm$ 0.19         &    $ld_2$               &         1.04 $\pm$ 0.23        \\
    \hline
    KIC~11442793~g (KOI~351.02) &                          &    KIC~11442793~c &                \\
    period (days)         & 210.606\,97  $\pm$ 0.000\,43   &    period (days)         &   8.719\,375 $\pm$ 0.000\,027 \\
    epoch (HJD - 2454833) & 147.036\,4   $\pm$ 0.001\,4    &    epoch (HJD - 2454833) & 139.568\,7   $\pm$ 0.002\,3   \\
    duration (h)          &      12.593  $\pm$ 0.045       &    duration (h)         &          4.41 $\pm$ 0.18       \\
    $a/R_s$               &       127.3  $\pm$ 4.1         &    $a/R_s$               &        16.0  $\pm$ 0.8        \\
    $a$ (AU)              &        0.71  $\pm$ 0.08        &    $a$ (AU)              &       0.089  $\pm$ 0.012      \\
    $R_p/R_s$             &      0.0615  $\pm$ 0.0011      &    $R_p/R_s$             &      0.0091  $\pm$ 0.0003      \\
    $R_p$ ($R_E$)         &         8.1  $\pm$ 0.8         &    $R_p$ ($R_E$)         &        1.19  $\pm$ 0.14        \\
    $b$                   &        0.45  $\pm$ 0.10        &    $b$                   &        0.09  $\pm$ 0.20       \\
    $i$ (deg)            &         89.80 $\pm$ 0.06        &    $i$ (deg)             &        89.68 $\pm$ 0.74       \\   
    $M^{1/3}/R_s$         &        0.84  $\pm$ 0.14         &    $M^{1/3}/R_s$          &       0.90  $\pm$ 0.16       \\
    $ld_1$               &         0.34 $\pm$ 0.10         &    $ld_1$               &         0.40 $\pm$ 0.20        \\
    $ld_2$               &         0.98 $\pm$ 0.10         &    $ld_2$               &         1.21 $\pm$ 0.26        \\
    \hline
    KIC~11442793~f &                                       &    KIC~11442793~b &                \\
    period (days)         & 124.914\,4   $\pm$ 0.001\,9    &    period (days)         &   7.008\,151 $\pm$ 0.000\,019 \\
    epoch (HJD - 2454833) & 254.704      $\pm$ 0.014       &    epoch (HJD - 2454833) & 137.690\,6   $\pm$ 0.001\,7   \\
    duration (h)          &       10.94  $\pm$ 0.25        &    duration (h)         &         3.99  $\pm$ 0.15       \\
    $a/R_s$               &        86.4  $\pm$ 9.7         &    $a/R_s$               &        13.2  $\pm$ 1.8        \\ %%15.3  $\pm$ 2.0        \\
    $a$ (AU)              &        0.48  $\pm$ 0.09        &    $a$ (AU)              &       0.074  $\pm$ 0.016      \\ %% 0.085  $\pm$ 0.018       \\
    $R_p/R_s$             &      0.0220  $\pm$ 0.0022      &    $R_p/R_s$             &      0.0100  $\pm$ 0.0005     \\ %% 0.0098  $\pm$ 0.0006      \\
    $R_p$ ($R_E$)         &        2.88  $\pm$ 0.52        &    $R_p$ ($R_E$)         &        1.31  $\pm$ 0.17        \\ %% 1.28  $\pm$ 0.19        \\
    $b$                   &        0.35  $\pm$ 0.40        &    $b$                   &        0.13  $\pm$ 0.32       \\ %% 0.12  $\pm$ 0.32       \\
    $i$ (deg)            &         89.77 $\pm$ 0.31        &    $i$ (deg)             &        89.4 $\pm$ 1.5         \\   %% 89.6 $\pm$ 1.3       \\   
    $M^{1/3}/R_s$         &        0.84  $\pm$ 0.20         &    $M^{1/3}/R_s$          &       0.85  $\pm$ 0.21        \\ %% 0.99  $\pm$ 0.24       \\
    $ld_1$               &        0.360 $\pm$ 0.068        &    $ld_1$               &        0.378 $\pm$ 0.060       \\
    $ld_2$               &         1.01 $\pm$ 0.18         &    $ld_2$               &         1.11 $\pm$ 0.20        \\
    \hline
    KIC~11442793~e        &                                & & \\
    period (days)         &  91.939\,13  $\pm$ 0.000\,73   & & \\
    epoch (HJD - 2454833) & 134.312\,7   $\pm$ 0.006\,3    & & \\
    duration (h)          &        9.71  $\pm$ 0.19        & & \\
    $a/R_s$               &        74.7  $\pm$ 4.3         & & \\
    $a$ (AU)              &        0.42  $\pm$ 0.06        & & \\
    $R_p/R_s$             &      0.0203  $\pm$ 0.0005      & & \\
    $R_p$ ($R_E$)         &        2.66  $\pm$ 0.29        & & \\
    $b$                   &        0.27  $\pm$ 0.22        & & \\
    $i$ (deg)             &        89.79 $\pm$ 0.19        & & \\
    $M^{1/3}/R_s$         &        0.87  $\pm$ 0.15        & & \\
    $ld_1$               &        0.360 $\pm$ 0.049        & & \\
    $ld_2$               &         1.05 $\pm$ 0.17         & & \\
    \hline
  \end{tabular}
\end{table}

\subsection{Analysis of the geometry of the transits}
\label{subsec:geometry}

One argument supporting the hypothesis that all these planet
candidates orbit the same star comes from the modeling of the
planetary parameters. 
The inclinations and stellar densities ($M^{1/3}/R_s$) shown in
Table~\ref{table:planets} were calculated independently for each
planet.
They are all compatible to each other and the density is compatible
with the value obtained independently for the stellar parameters in
Section~\ref{sec:star}.

\begin{figure}
  \centering
  \includegraphics[%
      width=0.9\linewidth,%
      height=0.5\textheight,%
      keepaspectratio]{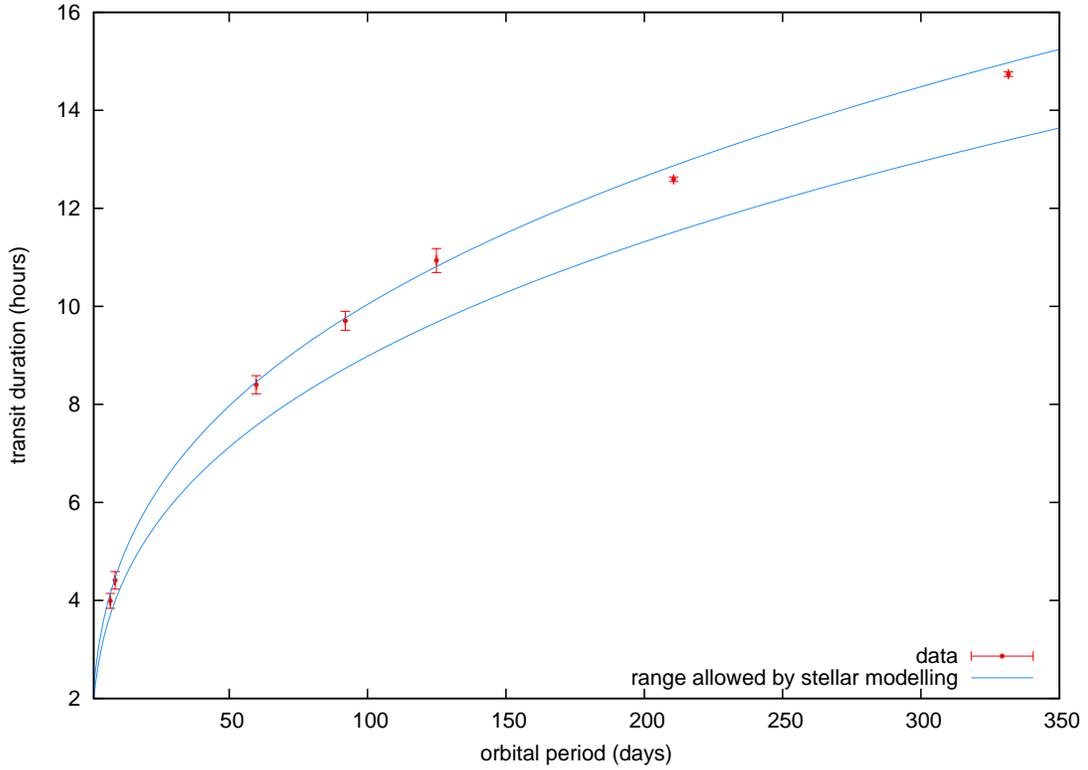}
  \caption{Transit duration of each planet as a function of their
    orbital period. 
    The observed values are compatible with the seven planets orbiting
    a star whose density is that given by the stellar and planetary
    parameter modelling on edge-on aligned orbits.
    The range allowed by the modelling of the stellar parameters is
    indicated with the continuous blue lines.
  }
  \label{figure:transitdurations}
\end{figure}

We can also provide another geometrical argument supporting the former
hypothesis using the measured durations and periods of the transiting
planets. 
These are obtained from a pure geometrical fit to the transits,
independently of the planetary modelling techniques. 
This argument has previously been used in the literature to support
the hypothesis that multiple candidate systems actually orbit the same
star\micitap{chaplin2013}.
Figure~\ref{figure:transitdurations} shows how the transit durations
distribute as a function of planetary periods. 
If all planets orbit the same star in circular, coplanar orbits, the
transit duration $D$ should relate to the orbital period $P$ through
Kepler's third law:
\begin{equation}
D = \frac{\alpha}{\pi} P^{1/3} \sqrt{ 1 - \left( \frac{\cos i}{\alpha}^2 P^{4/3} \right)},
\end{equation}
where $\alpha = ( 3 \pi/G/\rho_s)^{1/3}$, and $\rho_s$ is the density
of the star. 
If $D$ and $P$ are in days, the best fit to the data gives a value of
$\alpha = 0.23$ and $i = 90^\circ$, compatible with the values
obtained from the stellar and planetary modelling.
Note that the fit is not a physical solution, because all planetary
orbits do not need to be exactly coplanar. 
However, they are compatible with all planets orbiting the same star
in nearly edge-on aligned orbits, which supports our hypothesis that
all planets orbiting the same star.

%
%________________________________________________________________
\section{Transit timing variations}
\label{sec:ttv}

The analysis of the transit timing variations (TTVs) has proved to be
a versatile tool to confirm the planetary nature of transiting
candidates\micitap{ford2011}. 
Typically, TTVs have amplitudes of several minutes (with some
exceptional cases like KOI~142,\micitaalt{nesvorny2013a}, with an
amplitude of 12h) and typically periods one order of magnitude larger
than the orbital period of the planet involved\micitap{mazeh2013}.

\begin{figure}
  \centering
  \includegraphics[%
      width=0.9\linewidth,%
      height=0.5\textheight,%
      keepaspectratio]{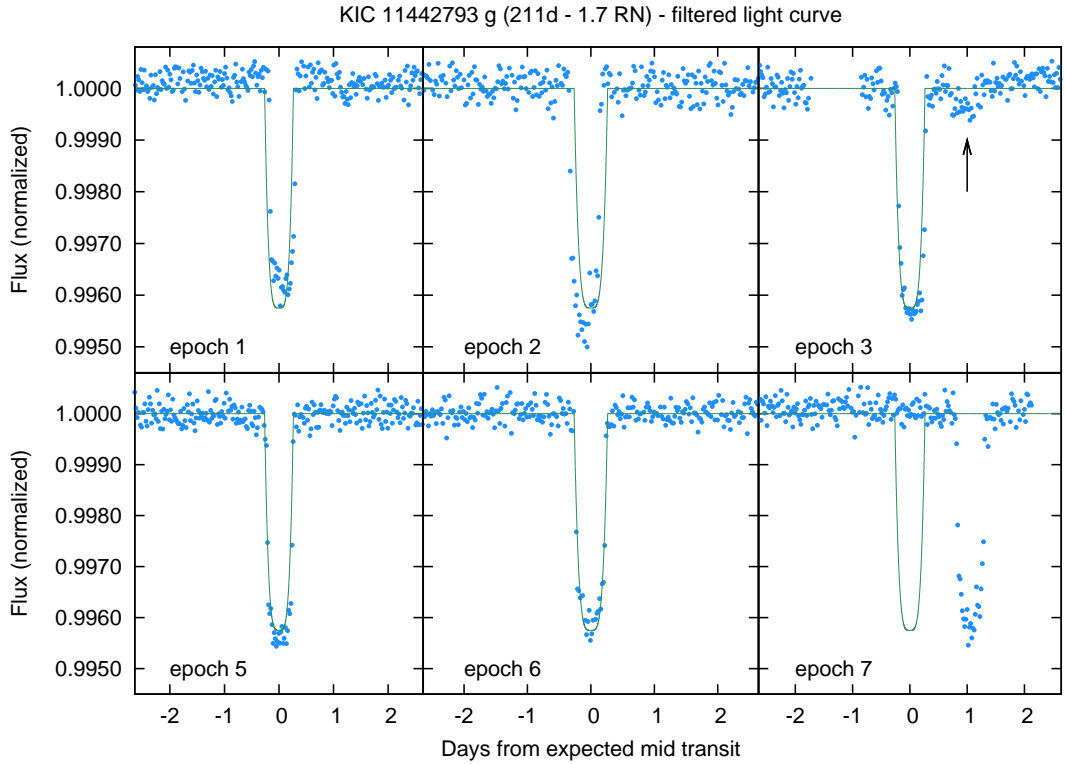}
  \caption{Individual observed transit events of planet g and the
    expected position of those transits assuming a constant period,
    marked with a line. 
    Note the irregularities in the transit depth and duration at
    epochs 1 and 2 and the displacement from the expected position of
    epoch 7.
    The additional transit like event marked with an arrow close to
    epoch 3 is discussed in the text.}
  \label{figure:transitsplanetg}
\end{figure}

\begin{figure}%[t]
  \centering
  \begin{minipage}[t]{0.48\textwidth}
    \begin{center}
      \includegraphics[%
        width=0.8\linewidth,%
        height=0.4\textheight,%
        keepaspectratio]{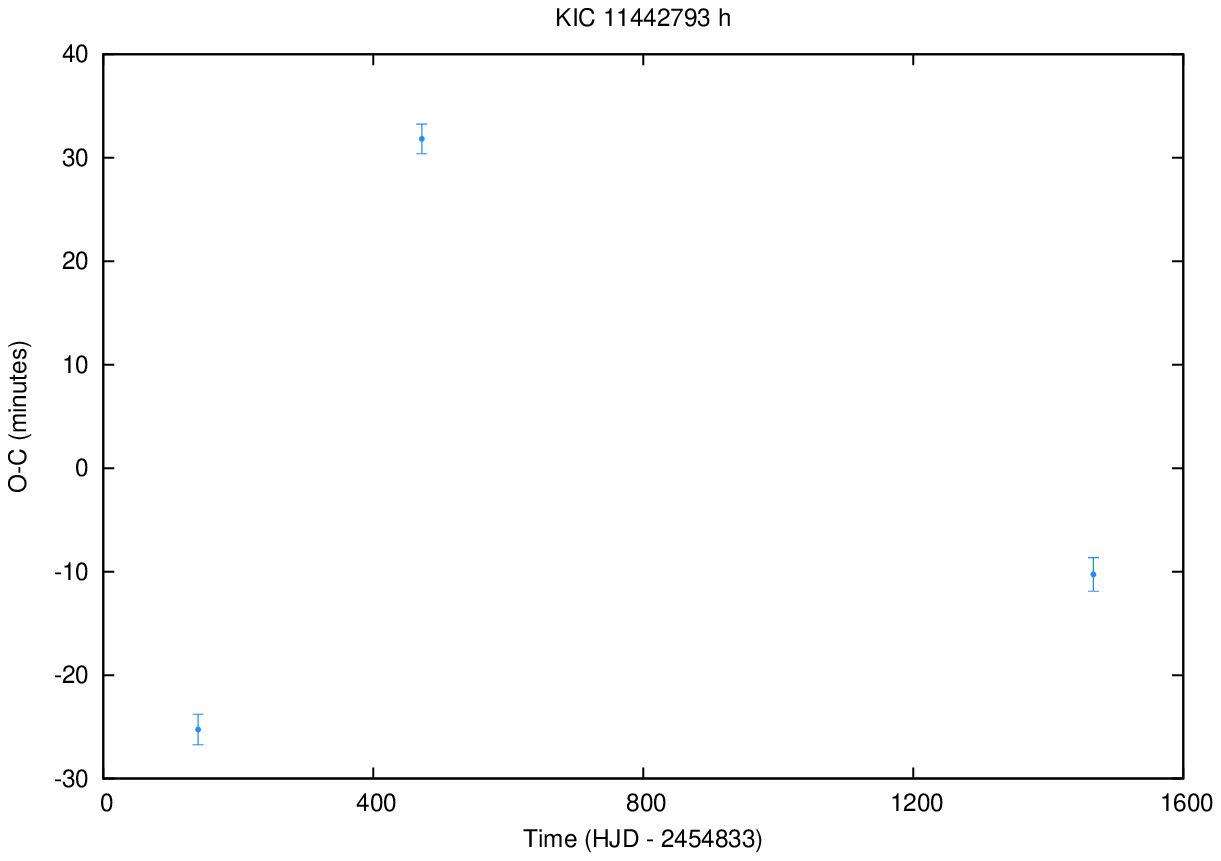}
      \includegraphics[%
        width=0.8\linewidth,%
        height=0.4\textheight,%
        keepaspectratio]{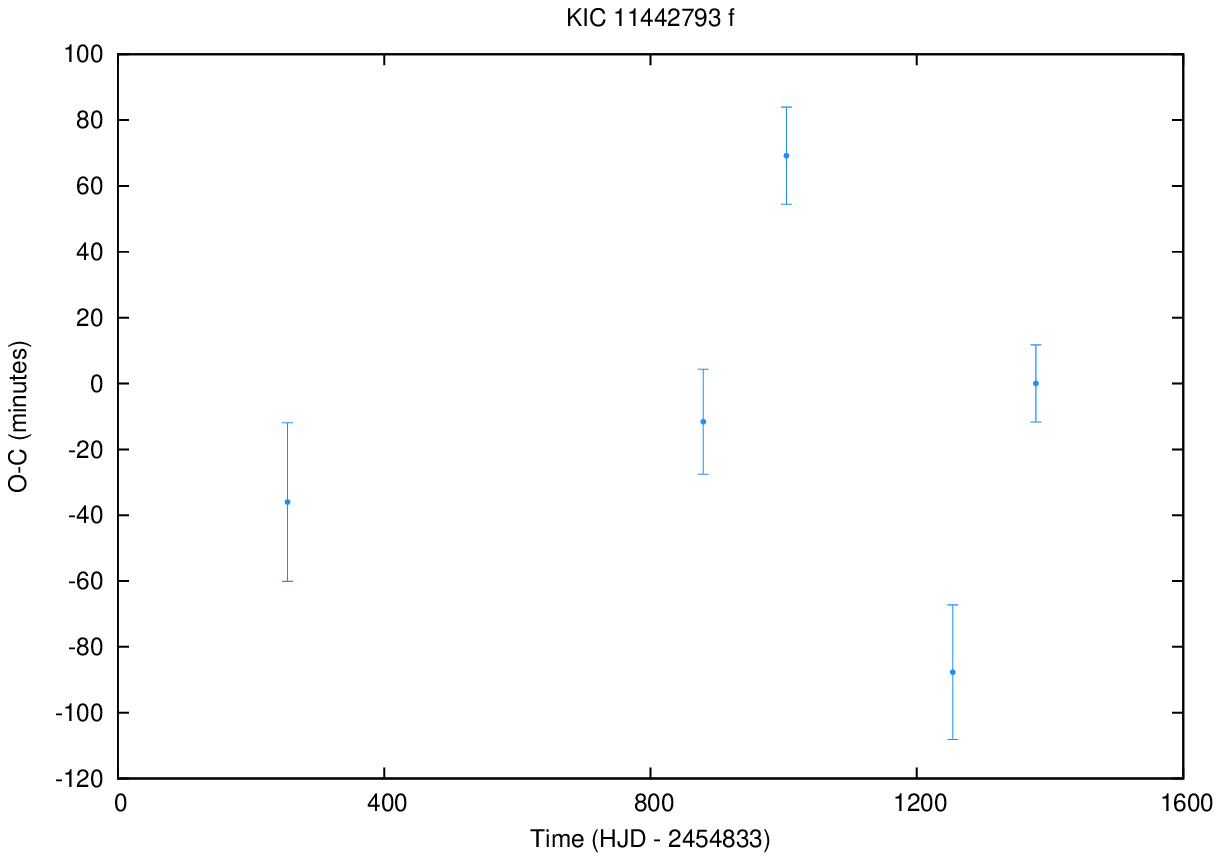}
      \includegraphics[%
        width=0.8\linewidth,%
        height=0.4\textheight,%
        keepaspectratio]{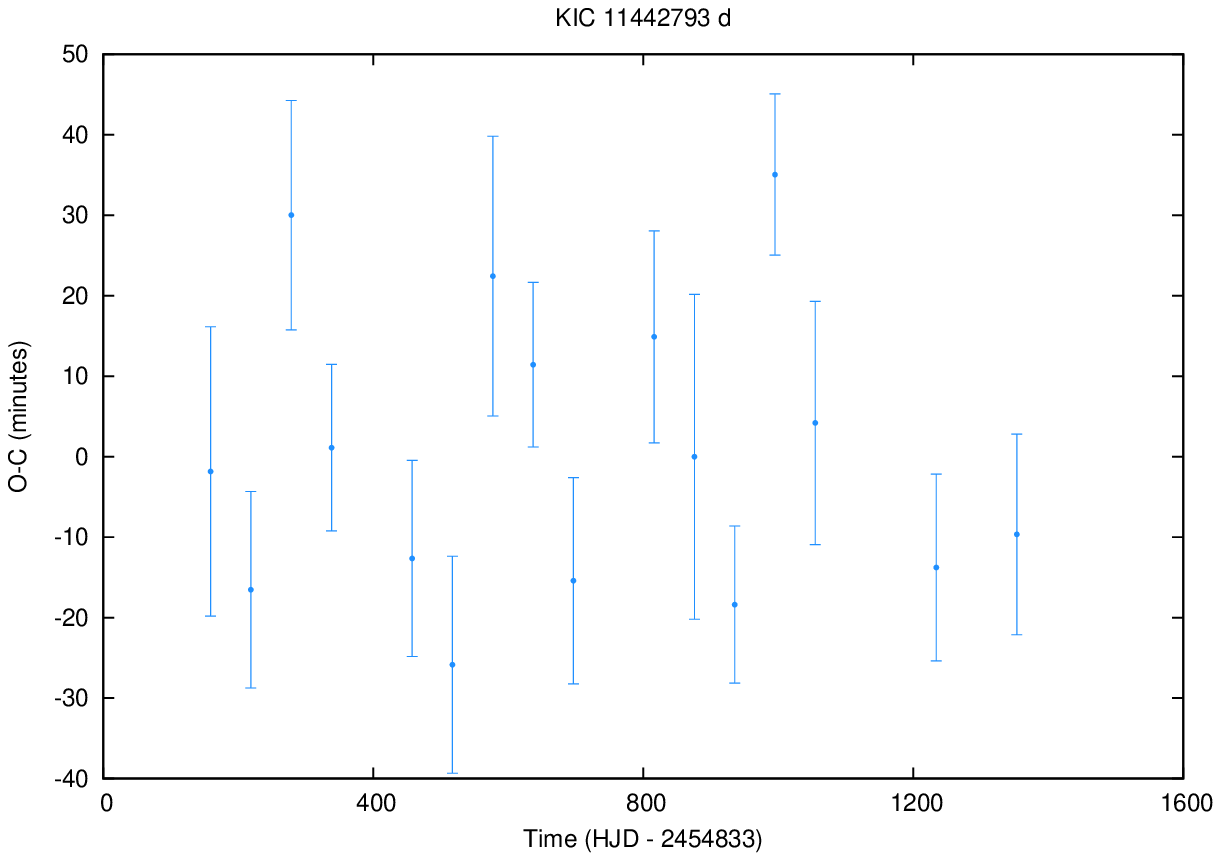}
      \includegraphics[%
        width=0.8\linewidth,%
        height=0.4\textheight,%
        keepaspectratio]{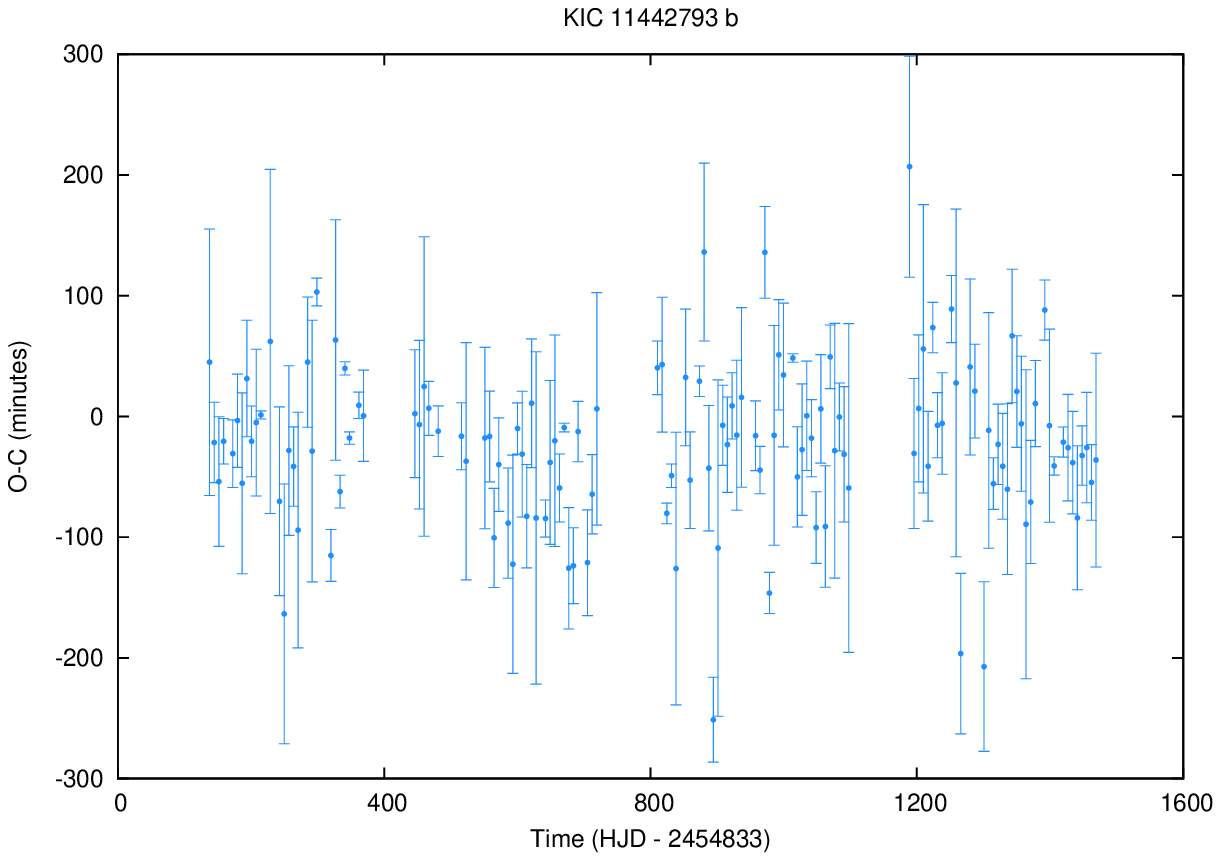}
    \end{center}
  \end{minipage}
  \begin{minipage}[t]{0.48\textwidth}
    \begin{center}
      \includegraphics[%
        width=0.8\linewidth,%
        height=0.4\textheight,%
        keepaspectratio]{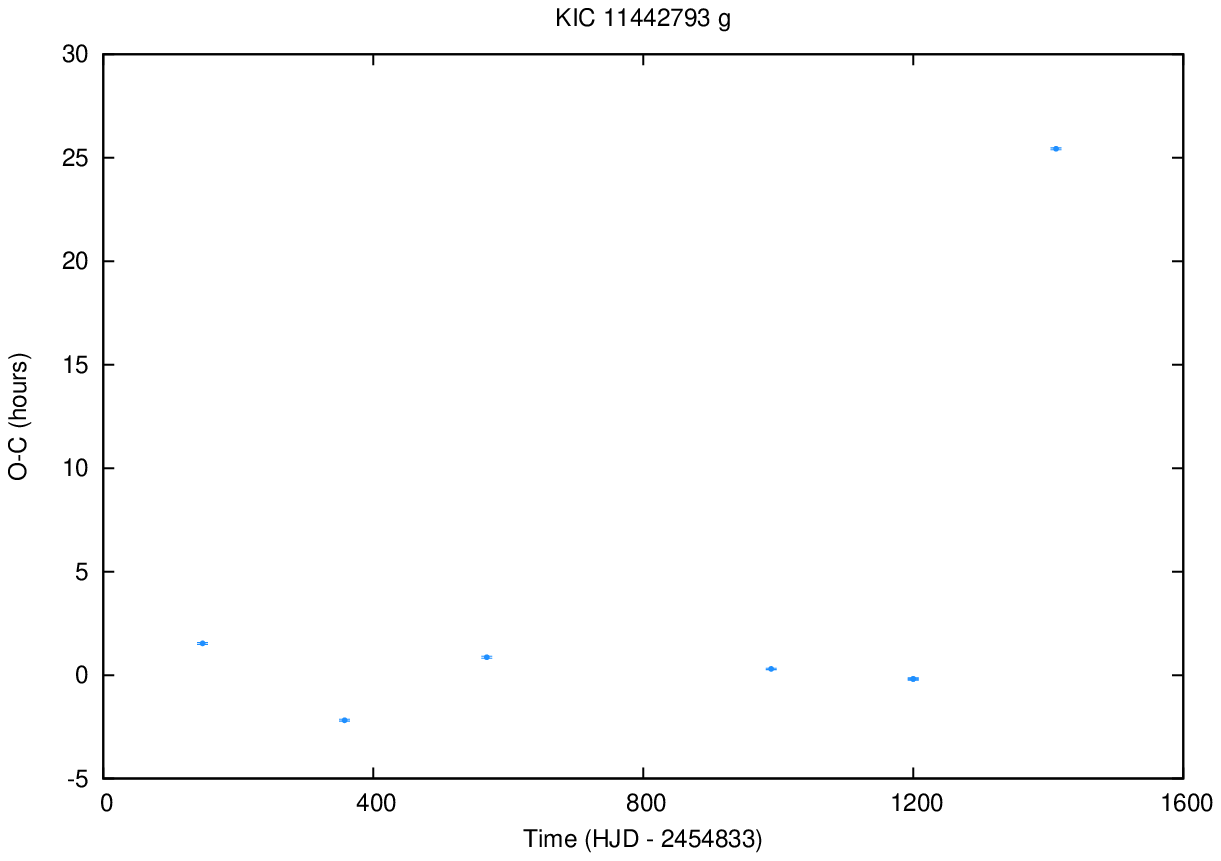}
      \includegraphics[%
        width=0.8\linewidth,%
        height=0.4\textheight,%
        keepaspectratio]{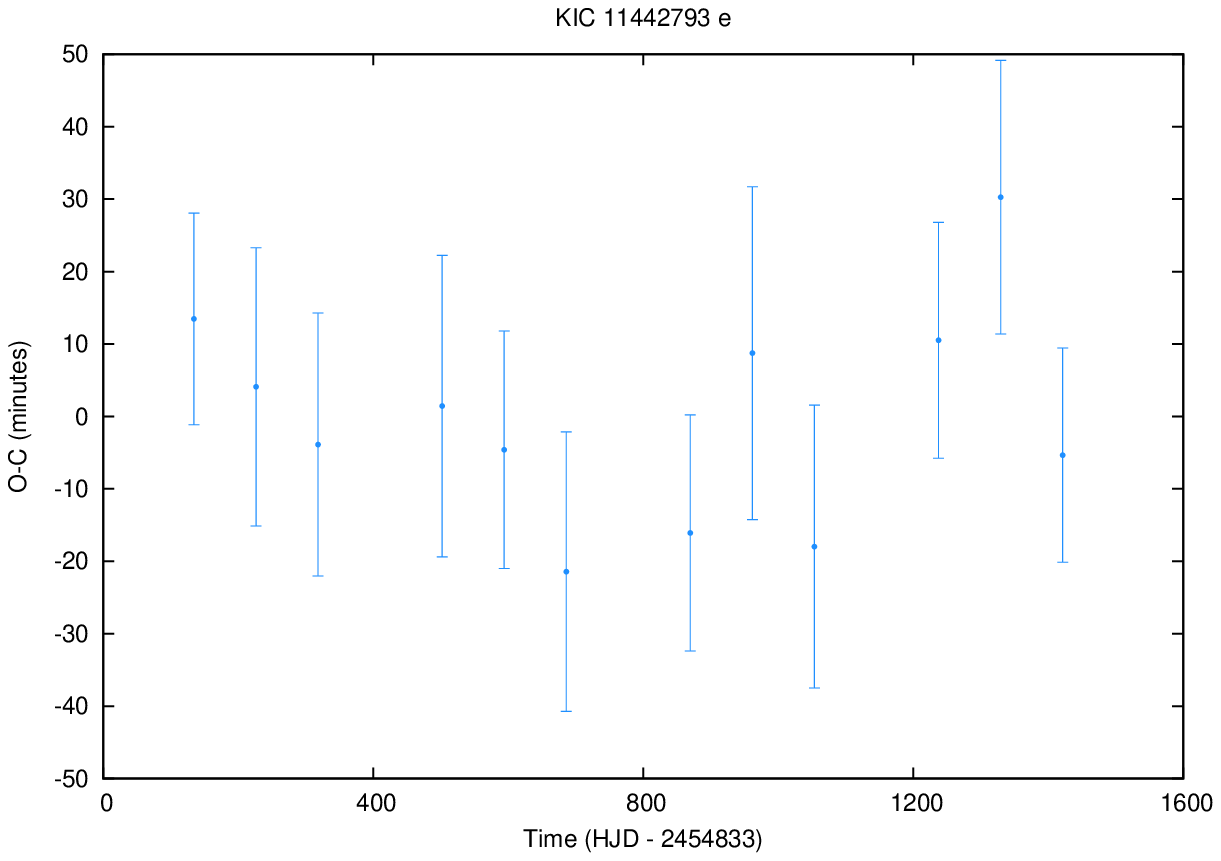}
      \includegraphics[%
        width=0.8\linewidth,%
        height=0.4\textheight,%
        keepaspectratio]{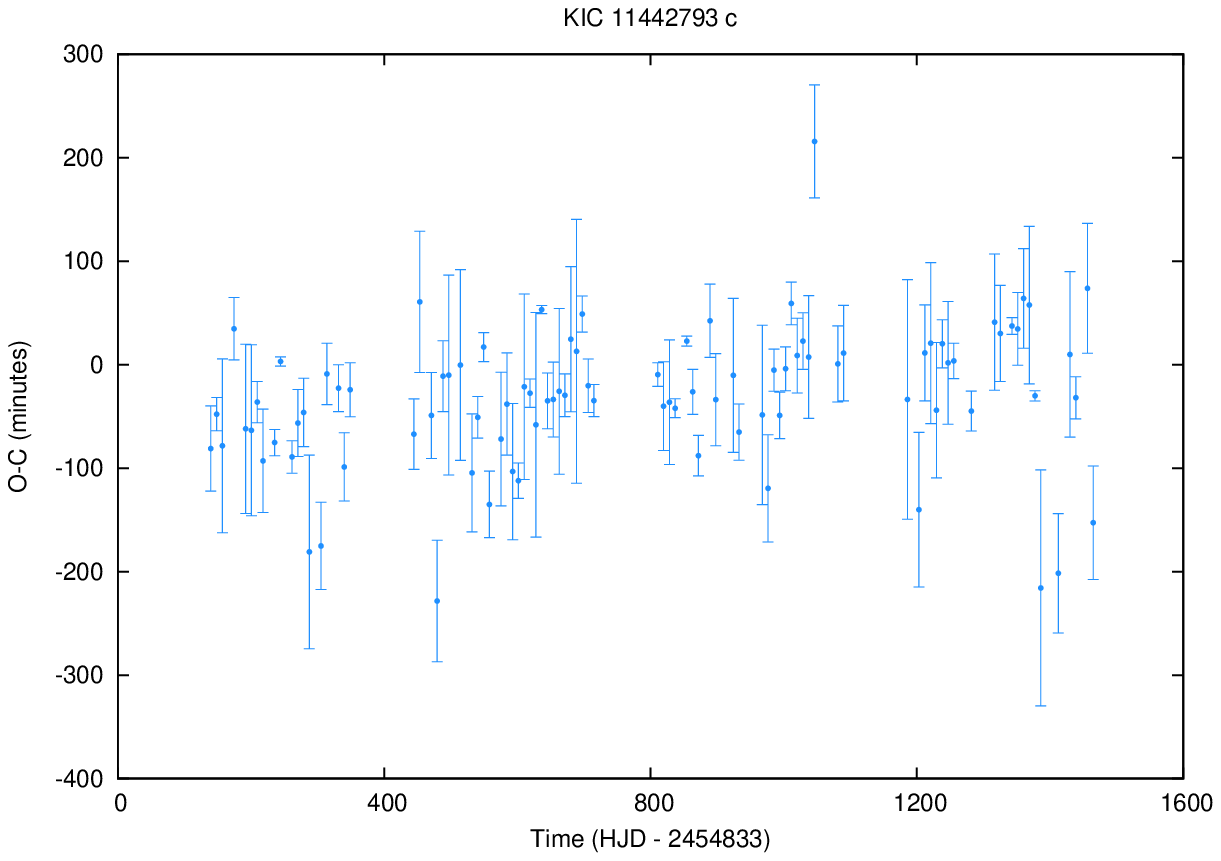}
    \end{center}
  \end{minipage}
  \caption{Transit timing variations of the different planets.
    Observed mid-times of planetary transits (O) minus calculated
    linear ephemeris (C) are plotted with 1$\sigma$ uncertainties.} 
  \label{figure:ttv}
\end{figure}

Figure~\ref{figure:transitsplanetg} shows the individual transits and
%Fig.~\ref{figure:ttvg} the O-C diagram for candidate g.
Fig.~\ref{figure:ttv} the O-C diagram for candidate g.
The transit corresponding to epoch 7 (epoch 1 being the value provided
in Table~\ref{table:planets}) has a displacement of 25.7 hours with
respect to its expected position.
This abrupt change is due to a change in the osculating orbital
elements produced by the gravitational interaction with other objects 
in the system, possibly candidate h (see Section~\ref{sec:dynamics}).
 Most surveys of TTVs expect discovering periodic modulations of
  the timing perturbations (see a derivation of the searched
  expression in\micitaalt{lithwick2012a} and the series of 
  papers \micitaalt{ford2011,ford2012a,steffen2012a,fabrycky2012a,ford2012b,steffen2012b,steffen2013,mazeh2013}).
  However, non-periodic, sudden changes of the orbital elements,
  corresponding to irregular behavior such as the one displayed by
  planet g, have been theoretically described (for example, though in
  a different context,\micitaalt{holman2005}), but we believe that we
  report an observational example for the first time. 

In addition to the change in the osculating elements, it is
interesting to discuss separately the other transit events recorded
for candidate g.
The depth and the duration of transit events 1, 2, and 3 changes
significantly. 
 One can speculate that the perturbations seen around these transits
are morphologically equivalent to those produced by a moon around the 
planet\micitap{sartoretti1999,kipping2013a}. 
This hypothesis is further discussed in Section~\ref{sec:moon}.
We do not have enough evidence to prove that these perturbations are
produced by a moon and until we have constraints on the planetary
masses we cannot assess the stability of moons around candidate g.  
% There are studies about the stability and the survival of moons around
% giant planets in multiple systems that affirm that although they might
% not be very common, such moons might exist\micitap{gong2013}. 
%% However, we cannot conclude with the current available data set in any
%% direction, either proving or disproving the presence of a moon.
We note, just for completion, that a moon could not be responsible in
any case for the abrupt change in the osculating orbital elements
displayed in transit event 7.
The amplitudes of the perturbations produced by moons are typically
only a few seconds\micitap{cabrera2008,kipping2009a,kipping2009b}. 
%%{IN THIS ORBIT IS IT ONLY SECONDS?}

The available data set for KIC~11442793 does not allow us to do an
unambiguous determination of the planetary masses from the analysis of
the TTVs. 
Candidates b and c are too small and too close to the detection limit
to measure any reliable TTV amplitude 
%{\bf (see Figures~\ref{figure:ttvc} and~\ref{figure:ttvb})}, which is
 (see Figure~\ref{figure:ttv}), which is
not unusual in the case of low-mass planets in compact systems (see
the case of CoRoT-7b\micitaalt{leger2009}).
The TTVs of candidates d and e are compatible with zero within the
limits of our current modelling
%{\bf (see Figures~\ref{figure:ttve} and~\ref{figure:ttvd})}.
 (see Figure~\ref{figure:ttv}).
There are only 5 full transits observed from the 9 expected transits
of candidate f due to some unfortunate coincidence of observing
interruptions with the expected transit positions.
However, there is a significant signal in the available O-C diagram,
which means that candidate f is interacting dynamically with other
objects in the system. % {\bf (see Figure~\ref{figure:ttv})}.% {\bf (see Figure~\ref{figure:ttvf})}. 

Candidate g shows 6 transits in the available data set (expected 7)
and candidate h shows 3 transits (expected 5), less than expected due
to the interruptions of the photometric record (duty cycle is 82\%). 
However, candidates g and h show both significant TTVs and also
transit duration variations, consequently we deduce that they are
interacting dynamically.
% {\bf (see Figures~\ref{figure:ttvh} and~\ref{figure:ttvg})}. 
% {\bf (see Figure~\ref{figure:ttv})}. 

%
%________________________________________________________________
\section{Dynamical study}
\label{sec:dynamics}

%
%................................................................
\subsection{Analysis with a numerical integrator}

We have done a stability analysis of the system with the orbital
dynamics integrator {\em Mercury}\micitap{chambers1999}.
The system is only stable if candidates g and h have masses below some
Jupiter units (typically, less than 5 Jupiter masses). 
Therefore, we conclude that g and h are planets because they interact
gravitationally and their long term dynamical stability is only
guaranteed if these bodies have planetary masses.

%% http://oklo.org/2010/06/23/a-second-laplace-resonance/
%% l_io - 3l_eu +2l_ga = 180º
%%  the closest Laplace resonance for planets f , e and d is: 0.000759116
%%  2*PI*(  -2 / f   3 / e   -1 / d )
%%  2*PI*(  -2 / 124.914   3 / 91.9391   -1 / 59.7367 )
%%  pr 2*pi*(1./per3-3./per4+2./per5)
%%  0.00075911605314055
%% Candidates d, e, and f are close to a resonant chain 2:3:4 and the
%% analysis of the mean motions reveal that they are probably in a
%% Laplace resonance (see Section~\ref{sec:dynamics}). 
%% Moreover, we know because of the significant TTV that at least
%% candidate f is interacting dynamically with other planets (possibly g
%% and h) in the system.
The {\em Mercury} numerical analysis of the planetary system reveals
that objects d, e, and f are in stable orbits only if those are very
circular (typically, less than 3\% for mass values of 10 Earth masses, 
representative of 2.5 Earth radii super-Earths) and planetary masses
(less than the mass of Jupiter). 
Therefore, we conclude that these three must also be planets.
Actually, the requirement of the circularity of their orbits implies
that, for the system to be stable, the mean motion resonance has to
play a role to guarantee the survival of the system. 

We did not see any sign that candidates b and c interact
dynamically with the other planets in the system because the low SNR
of the transit light curves.
%% However, the fact that their signal appears in a multiple transiting
%% system and that their periods are within 0.5\% of the 4:5 mean motion
%% resonance speaks in favor of these two objects belonging to the same
%% system. 
The {\em Mercury} numerical analysis reveals that their orbits are in
principle only stable if the objects have planetary masses.

%% Complementary to the numerical study, we have checked the mutual
%% orbital distances between planets in units of their mutual Hill sphere
%% radius (see below), a convenient method to assess the stability of
%% multiple systems in circular, coplanar orbits\micitap{lissauer2011b}. 
%% The closest pair is g-h, separated by 4.6 Hills radii. 
%% The neccessary separation for a pair to be stable is around 3.5{\bf{$R_{Hill}$}}.

%
%................................................................
\subsection{A first dynamical study}

We estimated the masses of the seven planets considering their sizes
and assuming representative mean densities for each planetary class
(gas giant, ice giants, large and small super-Earth) as follows: 
planet 
$m_h = 0.8 M_{\textrm{Jupiter}}$, 
$m_g = 0.7 M_{\textrm{Neptune}}$,  
planets $m_f \sim m_e \sim m_d \sim 10 M_{\textrm{Earth}}$ and planets 
$m_c \sim m_b \sim 3 M_{\textrm{Earth}}$. 
Given the the periods and the estimated semi-major axes we can
compute their separation in terms of Hill radii.
Using the formula given below\micitap{chambers1996}:
\begin{equation} 
H= \left( \frac{m_1+m_2}{3} \right)^{\frac{1}{3}} \frac{a_1+a_2}{2}
\end{equation} 
we get the following numbers for the separation of neighboring
planets in Hill radii:
\begin{equation} 
g-h: 5, h-f: 11, f-e: 5, e-d: 10, d-c: 47, \;\mathrm{and}\; c-b: 10.
\end{equation} 

This indicates the stability of the different subsystem given they are
moving in almost circular orbits. 
Especially the inner planets b, c, d, e and f are relative safe in
their orbits, which is evident from their Hill radii. 
It is interesting to note, that the innermost two planets are in a 4:5
mean motion resonance (MMR); the two massive outer ones are not far
from a 5:8 MMR. 
It is also worth to mention that the three planets d, e and f are close 
to the interesting Laplace resonance, which is known to happen for the
the motion of the three Galilean Moons of Jupiter (Io, Europa and
Ganymede) but also for the three Moons of Uranus (Miranda, Ariel and
Umbriel, e.g.\micitaalt{ferrazmello1979}): 
\begin{equation}
\frac{1}{P_d} - \frac{3}{P_e} + \frac{2}{P_f} \sim 0
\end{equation}  

Because the inner system consisting of super-earth planets is quite
stable we concentrate on the dynamics of the planets h and g.
The stability of this extrasolar planetary system seems to depend on
the stability of the orbits of these two outer gas giants, which may
have even eccentric orbits given the relative distance to the star. 
So we tried to find borders for stable motion of the two outer gas
giants using the results of long term integrations up to $10^7$
years.\footnote{As integration method we used the a high precision
  LIE-integrator with automatic step
  (e.g.\micitaalt{hanslmeier1984}).} 

\begin{figure}
\centering
\includegraphics[width=6cm,angle=270]{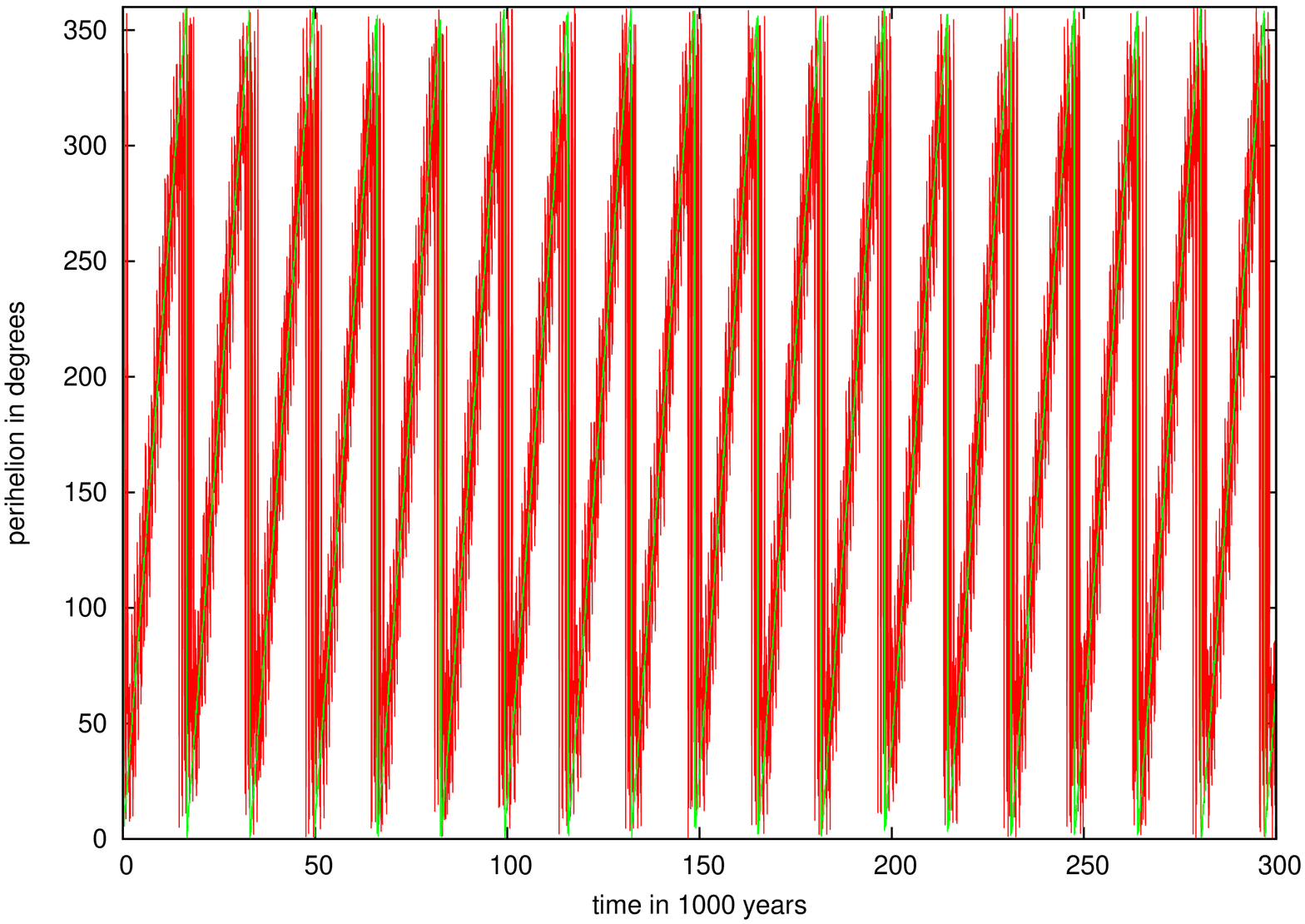}
\includegraphics[width=6cm,angle=270]{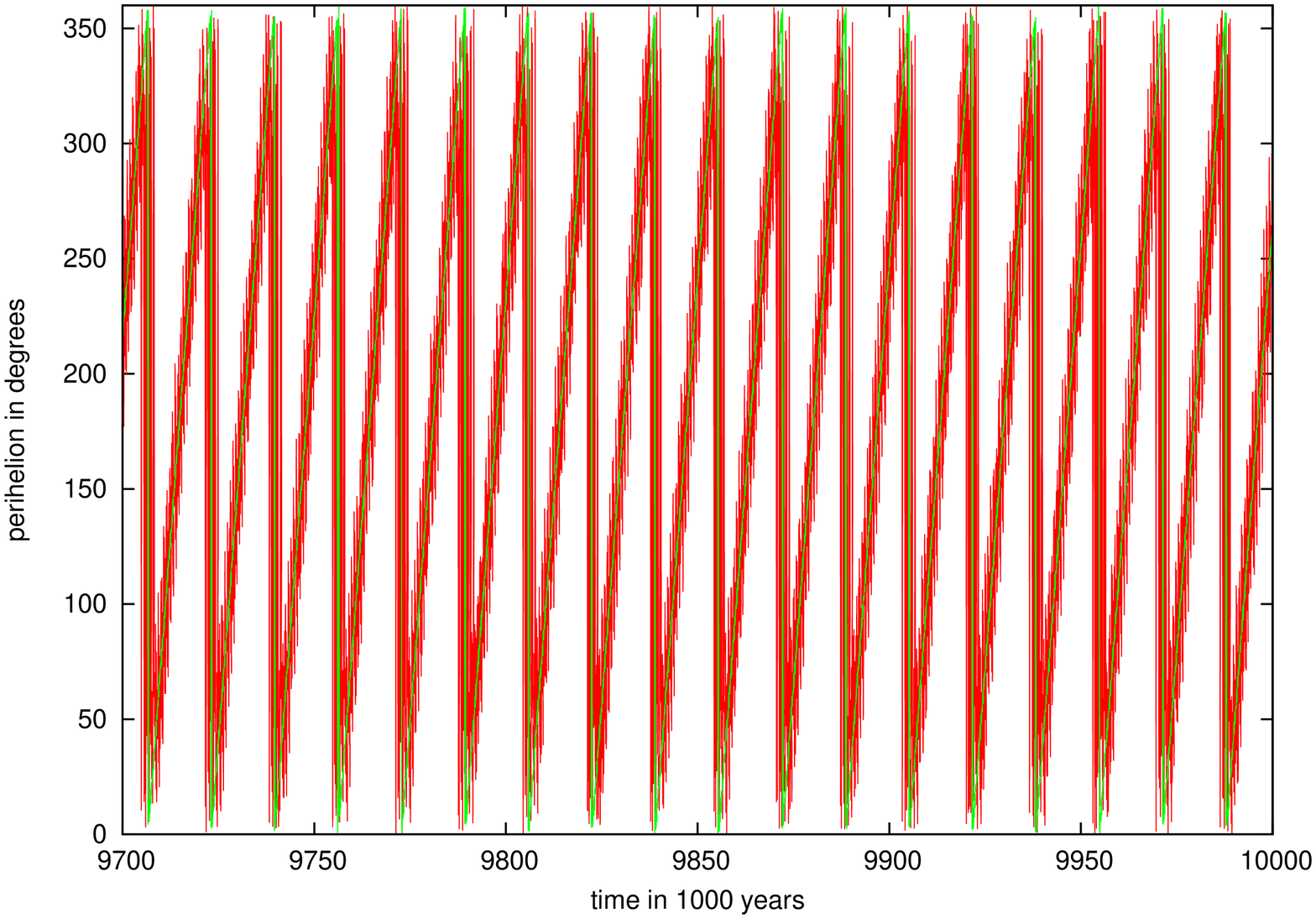}
\caption{Perihelion motion of the two outer planets h and g for
  initial conditions in the stable domain (see text). 
  Out of the whole integration time of $10^7$ years we show the first
  $3 \cdot 10^5$ (upper graph) and the last $3 \cdot 10^5$ years
  (lower graph). 
  The strong coupling in a 1:1 secular resonance of the perihelion
  motion with a period of around $17 \cdot 10^4$ is clearly visible.} 
\label{figure:omega}
\end{figure}

\begin{figure}
\centering
\includegraphics[width=4cm,angle=270]{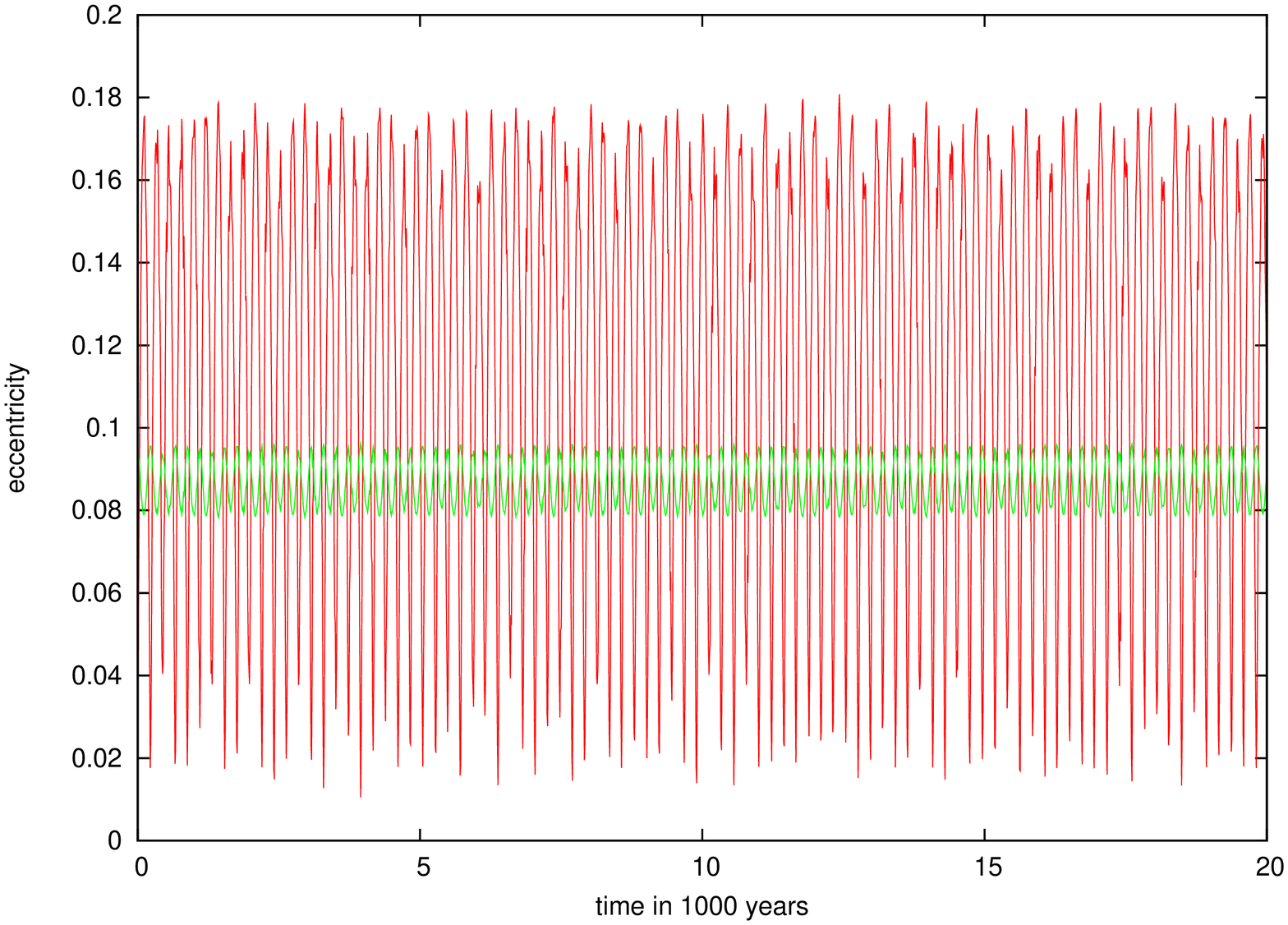}
\includegraphics[width=4cm,angle=270]{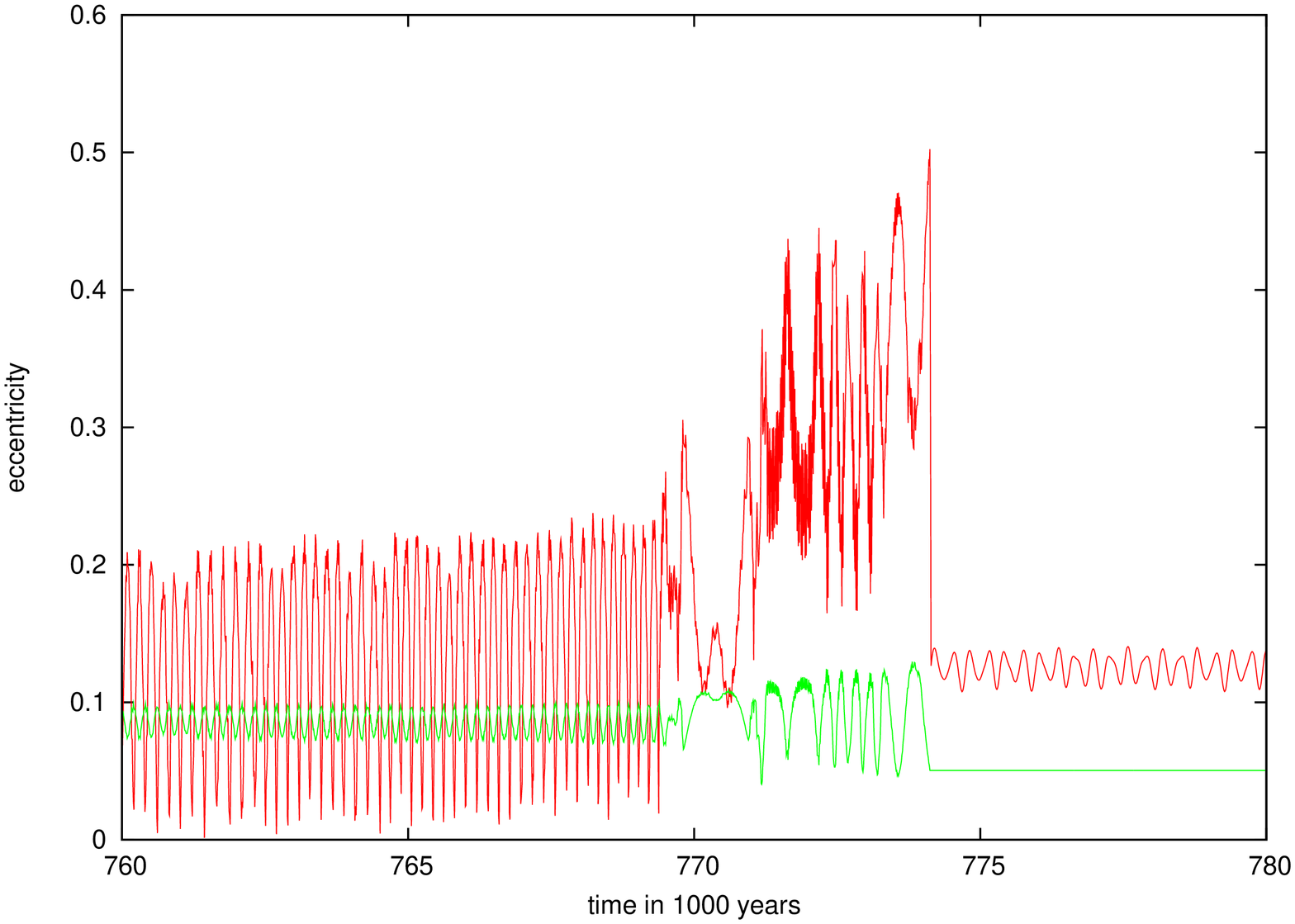}
\includegraphics[width=4cm,angle=270]{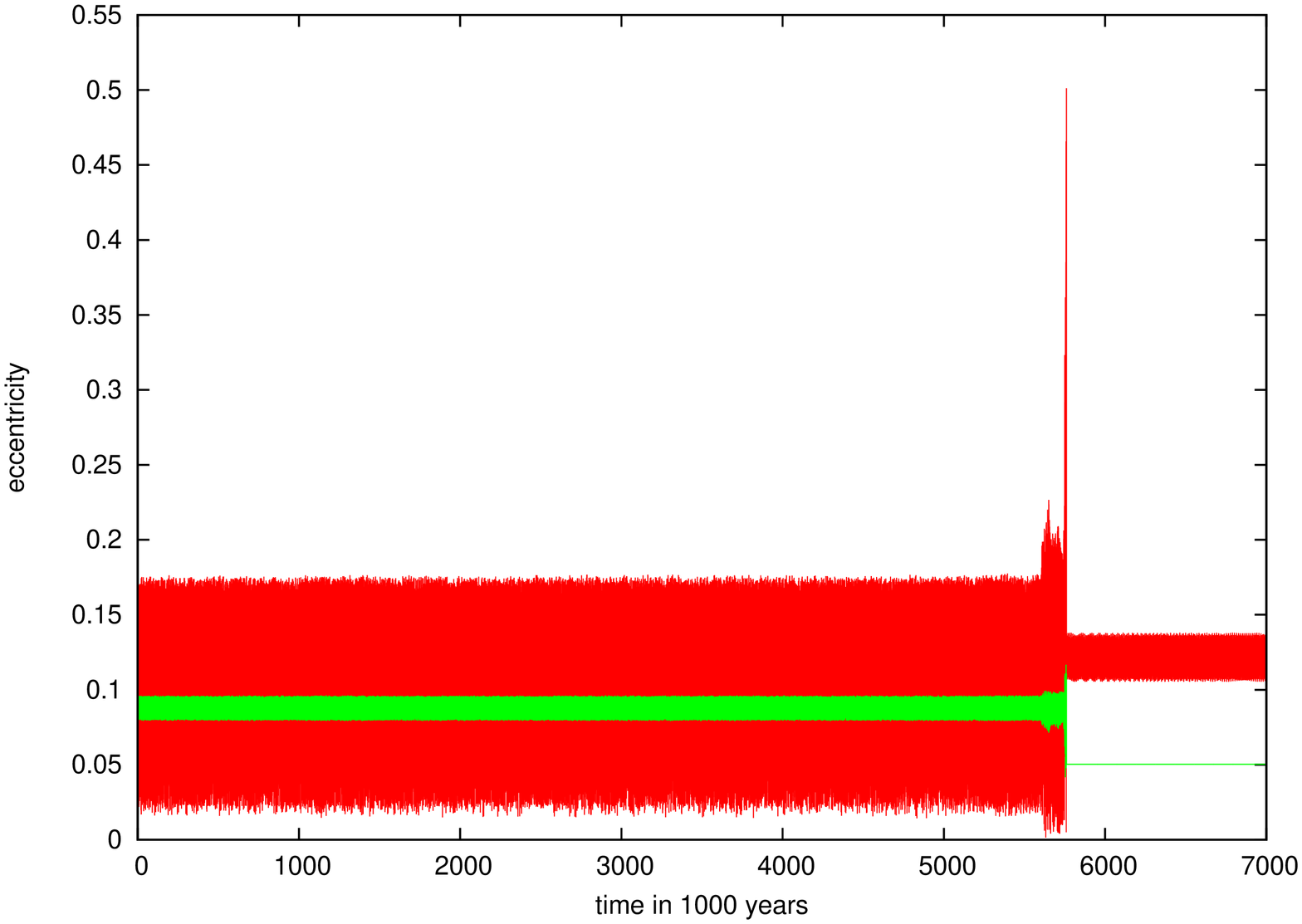}
\includegraphics[width=4cm,angle=270]{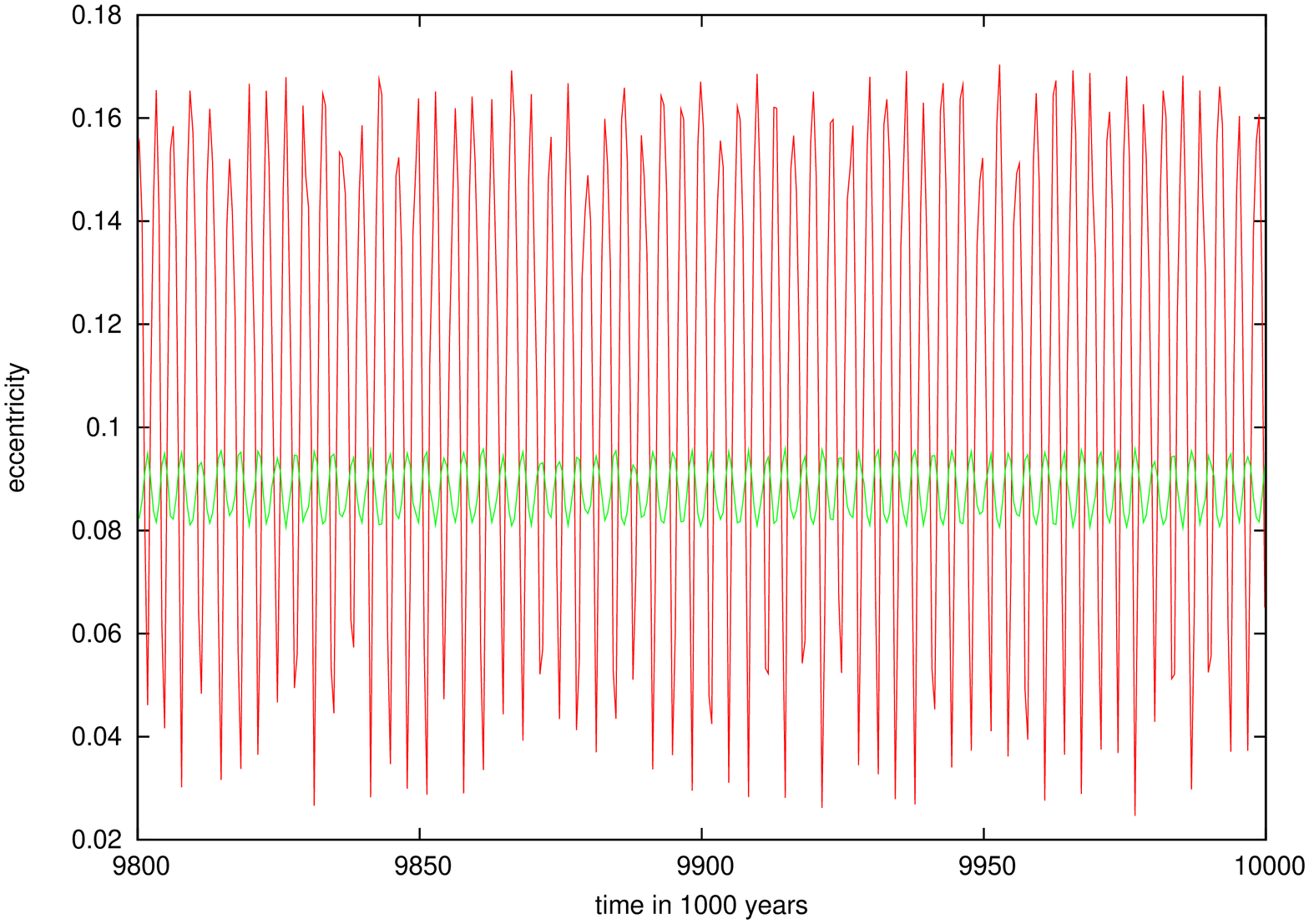}
\caption{Orbits close to the stability region: time evolution of the
  eccentricities of the orbits of planets h (green) and g
  (red). 
  Stable seeming orbit (upper left) which turns out to be  unstable
  after about $7.69 \cdot 10^5$ years (upper right). 
  Another close by orbit unstable after $5. \cdot 10^6$ years (lower
  left); another close by, but stable orbit is shown (lower right). 
  For details, see text.}
\label{figure:chaos}
\end{figure}

It turned out that inside the domain of motion for  
$e_h < 0.095$ and $e_g < 0.025$ the orbits of
the two outer planets are regular with only slight periodic changes in the
eccentricities (see Fig.~\ref{figure:chaos}, lower right graph).
The closeness to the 8:5 MMR is not destroying their stability;
an additional resonance appears for the motions of the
perihelia of  g and h. 
This secular resonance is depicted in Fig.\ref{figure:omega},
where the 1:1 resonance of the motion of $\omega_g$ and $\omega_h$ with
a period of about $1.7 \cdot 10^4$ years is visible. 

Close to the edge of the stable region we have an intermediate region
where stable and unstable orbits are very close to each other (see
Fig.~\ref{figure:chaos}).
In this domain we find the so-called sticky orbits - a well known
phenomenon of dynamical systems (e.g.\micitaalt{dvorak1998}): an orbit
there is 'sticked' to an invariant torus in phase space and then
escapes through a hole of the last KAM-torus.\footnote{KAM stands for
  Kolmogorov -- Arnold -- Moser.}
We show in the respective figure three such examples, where a small
shift in eccentricity of planet h ($\Delta e = 0.005$) causes such a
different dynamical behavior of an orbit.

\begin{figure}
\centering
\includegraphics[width=4cm,angle=270]{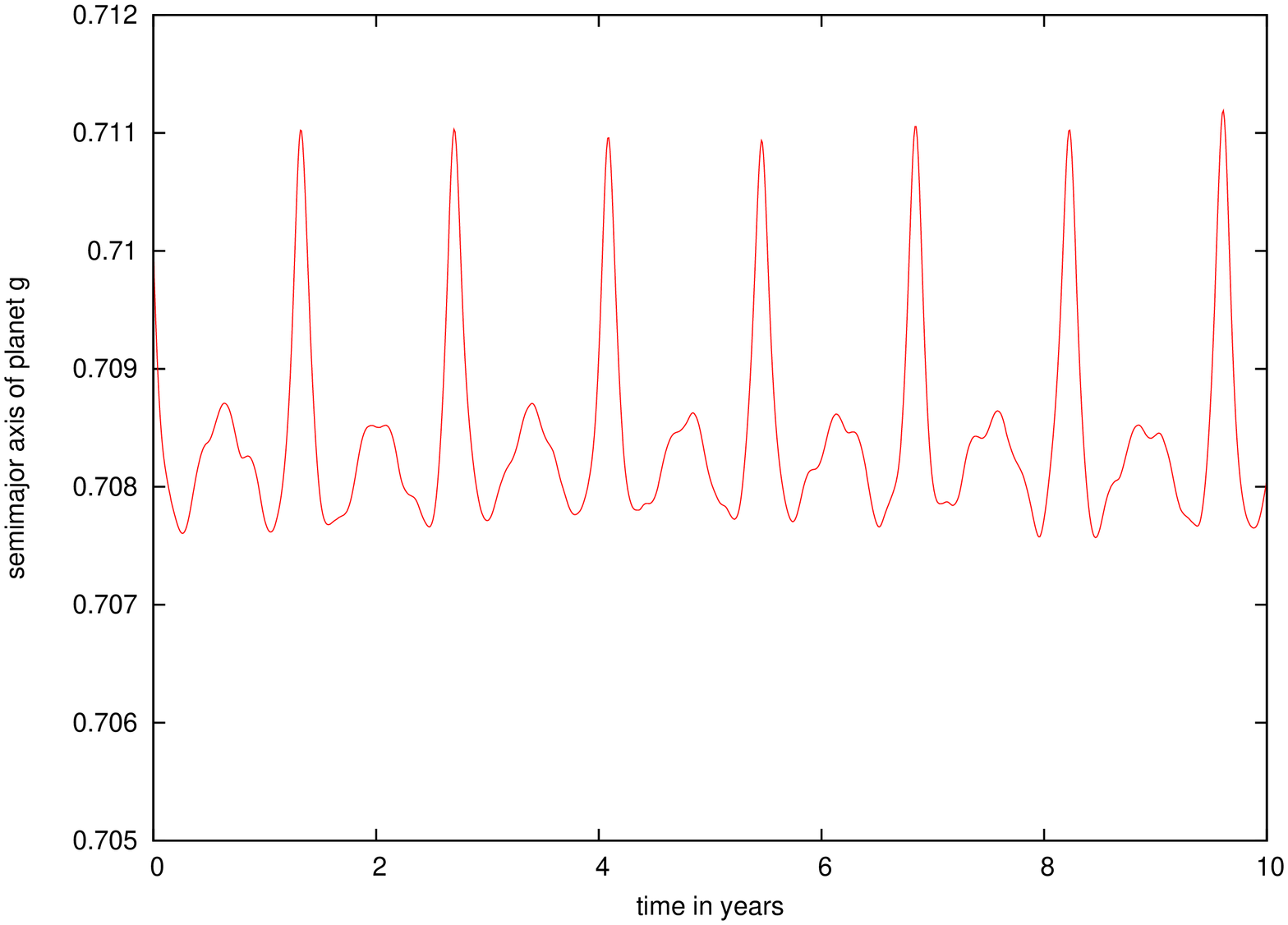}
\includegraphics[width=4cm,angle=270]{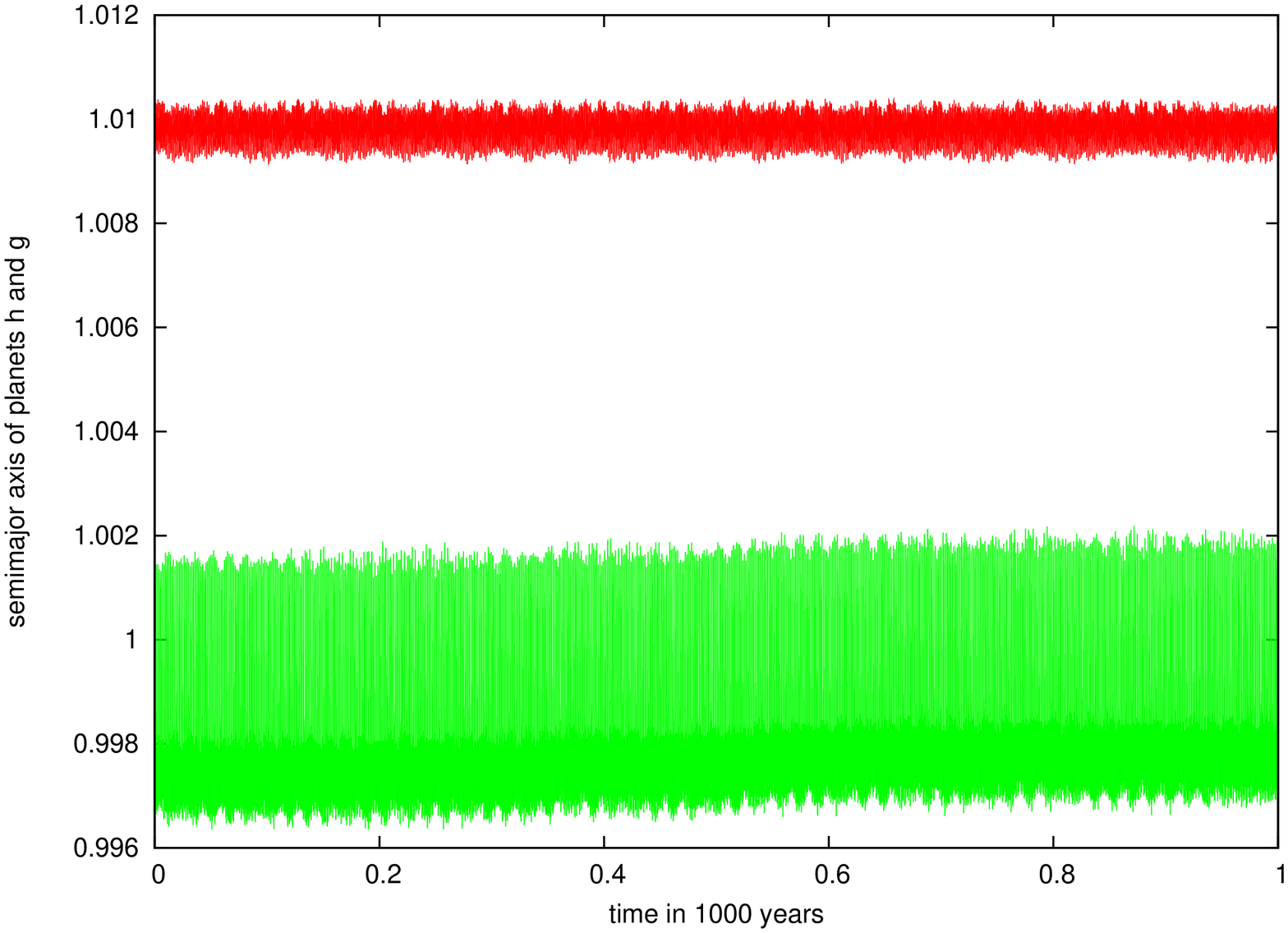}
\caption{Variation of the semi-major axis of planet g caused by the
  presence of planet h during 10 years (left graph). Variation of the
  semi-major axis of the outer gas giant caused by the inner gas giant
  g (red color) and vice versa on planet g (green color) during $10^3$
  years (right graph). 
  Note that the lower curve is normalized with respect to the
  semi-major axis of a=0.71 AU of planet g} 
\label{figure:ttvdvorak}
\end{figure}

We need to explain the large TTV for planet g: the answer is visible
from Fig.\ref{figure:ttvdvorak}, where one can see the relatively
large variations of the semi-major axis of this planet even for a time
scale of years. 
This change can lead to a change in the period which achieves values
up to a day from one transit to another one, comparable to the changes
observed in the \kepler~data.

But the system is quite more complex: because planet g is in 5:3 MMR 
with planet f and this one is in the formerly mentioned Laplace
resonance (with the planets e and d) the stability limit for the
eccentricities of all planets is very small.  
Integrating the 'complete' system\footnote{one can ignore the two
  innermost super-earth-planets - so we integrated the star plus the
  five outer planets} it is only stable well before the stability
limit mentioned above for the eccentricities of planets h and g: this
absolute limit for a stable system is $e < 0.001$ for all 5 outer
planets! 

We conclude from the preliminary dynamical study of this seven planet 
system that with the actual parameters determined it is quite close to
instability. 
Consequently the parameters like the masses and the semi-major axes
need some revision after a deeper dynamical study, out of the scope
of this paper. 
Even in our Solar System, where the orbital parameters are well
determined, the issue of the long term stability is debated
(e.g.\micitaalt{laskar1994,laskar2008}) and the influence of many
different resonances is complex.
We are currently working on that dynamical study (Dvorak et al. in
preparation).  

\section{KIC~11442793 in the context of other multiplanet systems}
\label{sec:discussion}

Models of planet formation include theories about planet-planet
scattering followed by tidal
circularization\micitap{rasio1996,lin1997,chatterjee2008,beauge2012a}. 
Another possible mechanism of planet formation builds planets at
relative large distances of the star and later these planets migrate
inwards through a
disk\micitap{goldreich1980,lin1996,ward1997,murray1998}.
The first mechanism does not likely form compact multiple systems such 
the ones observed by \kepler\micitap{batalha2013}, characterized by
being compact and by having low relative inclination
orbits\micitap{fang2012,tremaine2012}. 
Different mechanisms have been proposed to explain the origin of the
latter systems.
One promising possibility is in-situ formation, see for
example\micitap{chiang2013,chatterjee2013}, including the observed
feature that many of those systems have planets orbiting close to, but
not exactly at, mean motion
resonances\micitap{lithwick2012b,petrovich2013}. 

\begin{figure}
  \centering
  \includegraphics[%
      width=0.9\linewidth,%
      height=0.5\textheight,%
      keepaspectratio]{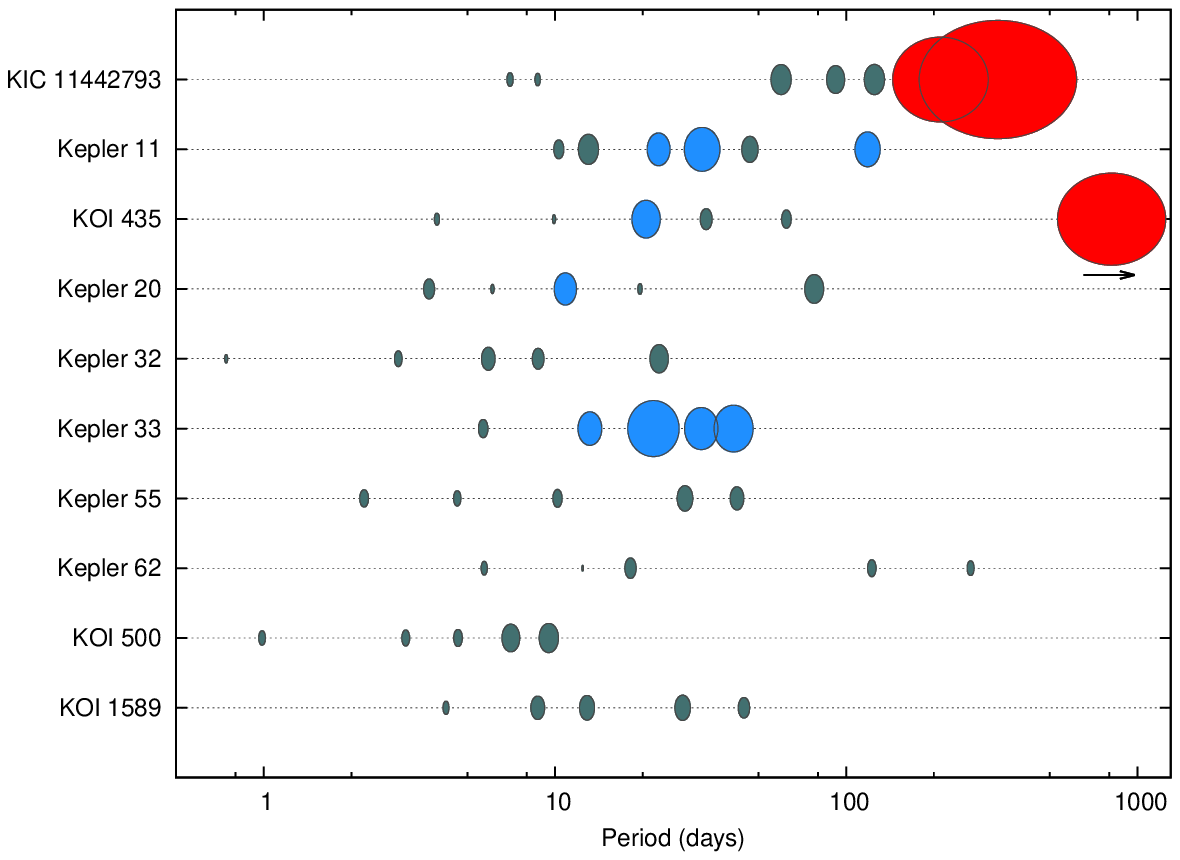}
  \caption{Comparison of different multiple systems. 
    Kepler-11\micitap{lissauer2011a}, 
    KOI-435\micitap{ofir2013a},
    Kepler-20\micitap{gautier2012,fressin2012a},
    Kepler-32\micitap{fabrycky2012a},
    Kepler-33\micitap{lissauer2012a},
    Kepler-55\micitap{steffen2013},
    Kepler-62\micitap{borucki2013},
    KOI-500\micitap{xie2013,wu2013},
    KOI-1589\micitap{xie2013,wu2013}.
    Color codes separate Earth and Super-Earth planets (up to 4 Earth
    radii, shown in green), Neptune-sized planets (between 4 and 8 Earth
    radii, shown in blue), and gas giants (larger than 8 Earth
    radii, shown in red).}
  \label{figure:systemcomparison}
\end{figure}

We show in Fig.~\ref{figure:systemcomparison} a schematic view of the
periods and relative sizes of 9 multiple transiting planetary systems
discovered by \kepler~with 5 transiting planets or more, together with
the planetary system reported in this paper.
There are also multiple systems discovered by radial velocity hosting
6 or more planets, like GJ~667C\micitap{angladaescude2013b},
HD~40307\micitap{tuomi2013}, or HD~10180\micitap{lovis2011}. 
But their orbital properties and even their existence is not as secure
as those of transiting candidates.
For example, consider the case of the system
GJ~581\micitap{hatzes2013b} or the discussion in the literature if
HD~10180 is orbited by six\micitap{feroz2011},
seven\micitap{lovis2011}, or even nine\micitap{tuomi2012} planets. 
Therefore, we limit ourselves in Fig.~\ref{figure:systemcomparison} to
the discussion of multiple transiting systems.
%% Among the systems shown, KIC~11442973 presented here is the only one
%% showing a clear hierarchy, like our Solar System, and the only one to
%% host a giant planet larger than 10 Earth radii.
%% {\bf Such systems are typically more difficult to form because giant
%%   planets tend to excite the excentricity of less masive planets
%%   during the migration processes, compromising the long term stability
%%   of the system (see, for example,\micitaalt{raymond2008}).}
  Among the systems shown, KIC~11442973 presented here is the only one
  showing a clear hierarchy, like our Solar System.
  Additionally, only KIC~11442973 and KOI~435 include a giant planet
  larger than 10 Earth radii. 
  Such systems are typically more difficult to form because giant
  planets tend to excite the eccentricity of less massive planets during
  the migration processes, compromising the long term stability of the
  system (see, for example,\micitaalt{raymond2008}).
Note that there are two additional known systems hosting
simultaneously super-Earths and gas giants, but these two systems
orbit M dwarfs, and only the second example is a compact system. 
GJ~676A\micitap{angladaescude2012a} hosts up to 4 planets, including
one super-Earth in a 3.6 days orbit and one 5 Jupiter masses planet in
a 1050 days orbit.
GJ~876\micitap{rivera2010} is also an M-dwarf hosting one super-Earth
of 6 Earth masses at 1.9 days orbital period, a 0.7 Jupiter masses
planet at 20 days, a 2.3 Jupiter masses planet at 61 days, and a 14
Earth masses planet in a 124 days orbit.
However, KIC~11442793 is a late F/early G solar-like star, hosting a
more complex system where dynamical interactions play an important
role in the long term stability of the system.

%% {\bf 
%%   could it be that this system is not dynamically {\em cold}?
%%   could it be that giant planets got expelled from this system,
%%   leaving only super-Earths?
%%   could it be that the Jupiter-like planet is sheperding the
%%   super-Earths? 
%%   citar dawson2012, tingley2011b para comparar TTVs
%% }

\subsection{About the possible existence of moons in the planetary system}
\label{sec:moon}

We have discussed in previous sections the possibility that
KIC~11442793g hosts a moon.
Figure~\ref{figure:transitsplanetg} shows that the transit epochs 1, 2
and 3 show features morphologically equivalent to an exomoon orbiting
the planet\micitap{sartoretti1999,szabo2006,kipping2011b}.
However, considering the distance between the transit epoch 3 and the
moon-like event marked with an arrow in
Figure~\ref{figure:transitsplanetg}, the estimated projected distance
between the planet and the exomoon candidate would be orbiting close
to the Hill radius of the planet, which is too far away to guarantee
the long term stability of the satellite, usually limited to a
distance of one third\micitap{barnes2002} to one
half\micitap{domingos2006} of the planetary Hill sphere.
With the current data set, we cannot exclude that the event marked
with an arrow in Figure~\ref{figure:transitsplanetg} is caused by
instrumental residuals.
However, the distorted shape features of transits 1 and 2 cannot be
explained simply by the impact of stellar activity and their origin
remains unclear.
Space surveys have regularly been used to rule out the presence of
moons around extrasolar planets\micitap{pont2007,deeg2010}.
So far, the most extensive search for exomoons\micitap{kipping2012c}
has taken the advantage of the simultaneous change in the transit
timing and transit duration changes produced by the hypothetical
satellites\micitap{kipping2009a,kipping2009b}.
However, until now only negative results have been
reported\micitap{kipping2013a,kipping2013b}. 
A possible reason for this lack of success is that searches have been
limited to isolated, typically non-giant planets.
However, if these systems are formed by planet-planet scattering, they
are unlikely to maintain the moons during their formation
process\micitap{gong2013}.
In turn, migration tends to remove moons from planetary
systems\micitap{namouni2010}.
Therefore, in-situ formed compact systems could be more prone to host
exomoons in long timescales. 

\section{Summary}
\label{sec:summary}

We report the discovery of a planetary system with seven transiting
planets with orbital periods in the range from 7 to 330 days (0.074 to
1.01 AU).
The system is hierarchical, the two innermost planets have sizes close
to Earth and their period ratio is within 0.5\% of the 4:5 mean motion
resonance. 
The three following planets are super-Earths with sizes between 2 and
3 Earth radii whose periods are close to a 2:3:4 chain. 
From the observational data set we cannot determine their masses or
the value of their mean longitudes, but the ratio of their mean
motions is close to a Laplace resonance. 
The outermost planets are two gas giants at distances of 0.7 and 1.0
AU.
There are other systems of super-Earths, discovered either by radial
velocity or by transit,which show some similarities, for example
GJ~876\micitap{rivera2010} or KOI~152\micitap{wang2012b}, 
but these systems only contain super-Earths, while KIC~11442793
is a hierarchical system.
As a singularity among the other multiple systems found by \kepler~or
radial velocity, KIC~11442793 contains a gas giant planet similar to
Jupiter orbiting at 1 AU. 
%% This distribution of sizes seems to be common among Kepler
%% candidates\micitap{ciardi2013}. 
Systems with super-Earths close to a Laplace resonance are also
believed to be frequent\micitap{chiang2013}, but this particular
system poses new challenges due to the presence of the gas giants g
and h, which seem to have the most intense gravitational
interaction measured among extrasolar planets so far (25.7 h of change
in the ephemeris). 
If \kepler~cannot continue the follow up of this
system\micitap{cowen2013}, the follow-up of the Earth and super-Earth
planets of this system will be challenging in the near future, as they
are beyond reach for CHEOPS\micitap{broeg2013} or
TESS\micitap{ricker2010}.
Only PLATO\micitap{rauer2011b} will be able to study in detail their
evolution. 
However, the gas giants g and f produce 0.5\% and 0.8\% transits,
which should be observable from ground, which makes of this system an
attractive target for future follow-up studies.

\acknowledgments
We are grateful to {\'E}.~B{\'a}lint, Ph.~von~Paris and M.~Godolt for
useful discussions concerning this paper.
This paper includes data collected by the \kepler~mission. Funding for
the \kepler~mission is provided by the NASA Science Mission
directorate. 
Some/all of the data presented in this paper were obtained from the
Mikulski Archive for Space Telescopes (MAST). STScI is operated by the
Association of Universities for Research in Astronomy, Inc., under
NASA contract NAS5-26555. Support for MAST for non-HST data is
provided by the NASA Office of Space Science via grant NNX09AF08G and
by other grants and contracts.

\bibliographystyle{apj}
\bibliography{bibl}

\begin{thebibliography}{}
\expandafter\ifx\csname natexlab\endcsname\relax\def\natexlab#1{#1}\fi

\bibitem[{{Alapini} \& {Aigrain}(2009)}]{alapini2009}
{Alapini}, A., \& {Aigrain}, S. 2009, \mnras, 397, 1591

\bibitem[{{Anglada-Escud{\'e}} \& {Tuomi}(2012)}]{angladaescude2012a}
{Anglada-Escud{\'e}}, G., \& {Tuomi}, M. 2012, \aap, 548, A58

\bibitem[{{Anglada-Escud{\'e}} {et~al.}(2013){Anglada-Escud{\'e}}, {Tuomi},
  {Gerlach}, {Barnes}, {Heller}, {Jenkins}, {Wende}, {Vogt}, {Butler},
  {Reiners}, \& {Jones}}]{angladaescude2013b}
{Anglada-Escud{\'e}}, G., {Tuomi}, M., {Gerlach}, E., {et~al.} 2013, \aap, 556,
  A126

\bibitem[{{Baglin} {et~al.}(2006){Baglin}, {Auvergne}, {Boisnard}, {Lam-Trong},
  {Barge}, {Catala}, {Deleuil}, {Michel}, \& {Weiss}}]{baglin2006}
{Baglin}, A., {Auvergne}, M., {Boisnard}, L., {et~al.} 2006, in COSPAR, Plenary
  Meeting, Vol.~36, 36th COSPAR Scientific Assembly, 3749

\bibitem[{{Barnes} \& {O'Brien}(2002)}]{barnes2002}
{Barnes}, J.~W., \& {O'Brien}, D.~P. 2002, \apj, 575, 1087

\bibitem[{{Batalha} {et~al.}(2010){Batalha}, {Rowe}, {Gilliland}, {Jenkins},
  {Caldwell}, {Borucki}, {Koch}, {Lissauer}, {Dunham}, {Gautier}, {Howell},
  {Latham}, {Marcy}, \& {Prsa}}]{batalha2010b}
{Batalha}, N.~M., {Rowe}, J.~F., {Gilliland}, R.~L., {et~al.} 2010, \apjl, 713,
  L103

\bibitem[{{Batalha} {et~al.}(2013){Batalha}, {Rowe}, {Bryson}, {Barclay},
  {Burke}, {Caldwell}, {Christiansen}, {Mullally}, {Thompson}, {Brown},
  {Dupree}, {Fabrycky}, {Ford}, {Fortney}, {Gilliland}, {Isaacson}, {Latham},
  {Marcy}, {Quinn}, {Ragozzine}, {Shporer}, {Borucki}, {Ciardi}, {Gautier},
  {Haas}, {Jenkins}, {Koch}, {Lissauer}, {Rapin}, {Basri}, {Boss}, {Buchhave},
  {Carter}, {Charbonneau}, {Christensen-Dalsgaard}, {Clarke}, {Cochran},
  {Demory}, {Desert}, {Devore}, {Doyle}, {Esquerdo}, {Everett}, {Fressin},
  {Geary}, {Girouard}, {Gould}, {Hall}, {Holman}, {Howard}, {Howell},
  {Ibrahim}, {Kinemuchi}, {Kjeldsen}, {Klaus}, {Li}, {Lucas}, {Meibom},
  {Morris}, {Pr{\v s}a}, {Quintana}, {Sanderfer}, {Sasselov}, {Seader},
  {Smith}, {Steffen}, {Still}, {Stumpe}, {Tarter}, {Tenenbaum}, {Torres},
  {Twicken}, {Uddin}, {Van Cleve}, {Walkowicz}, \& {Welsh}}]{batalha2013}
{Batalha}, N.~M., {Rowe}, J.~F., {Bryson}, S.~T., {et~al.} 2013, \apjs, 204, 24

\bibitem[{{Beaug{\'e}} \& {Nesvorn{\'y}}(2012)}]{beauge2012a}
{Beaug{\'e}}, C., \& {Nesvorn{\'y}}, D. 2012, \apj, 751, 119

\bibitem[{{Borucki} {et~al.}(2010){Borucki}, {Koch}, {Basri}, {Batalha},
  {Brown}, {Caldwell}, {Caldwell}, {Christensen-Dalsgaard}, {Cochran},
  {DeVore}, {Dunham}, {Dupree}, {Gautier}, {Geary}, {Gilliland}, {Gould},
  {Howell}, {Jenkins}, {Kondo}, {Latham}, {Marcy}, {Meibom}, {Kjeldsen},
  {Lissauer}, {Monet}, {Morrison}, {Sasselov}, {Tarter}, {Boss}, {Brownlee},
  {Owen}, {Buzasi}, {Charbonneau}, {Doyle}, {Fortney}, {Ford}, {Holman},
  {Seager}, {Steffen}, {Welsh}, {Rowe}, {Anderson}, {Buchhave}, {Ciardi},
  {Walkowicz}, {Sherry}, {Horch}, {Isaacson}, {Everett}, {Fischer}, {Torres},
  {Johnson}, {Endl}, {MacQueen}, {Bryson}, {Dotson}, {Haas}, {Kolodziejczak},
  {Van Cleve}, {Chandrasekaran}, {Twicken}, {Quintana}, {Clarke}, {Allen},
  {Li}, {Wu}, {Tenenbaum}, {Verner}, {Bruhweiler}, {Barnes}, \&
  {Prsa}}]{borucki2010a}
{Borucki}, W.~J., {Koch}, D., {Basri}, G., {et~al.} 2010, Science, 327, 977

\bibitem[{{Borucki} {et~al.}(2013){Borucki}, {Agol}, {Fressin}, {Kaltenegger},
  {Rowe}, {Isaacson}, {Fischer}, {Batalha}, {Lissauer}, {Marcy}, {Fabrycky},
  {D{\'e}sert}, {Bryson}, {Barclay}, {Bastien}, {Boss}, {Brugamyer},
  {Buchhave}, {Burke}, {Caldwell}, {Carter}, {Charbonneau}, {Crepp},
  {Christensen-Dalsgaard}, {Christiansen}, {Ciardi}, {Cochran}, {DeVore},
  {Doyle}, {Dupree}, {Endl}, {Everett}, {Ford}, {Fortney}, {Gautier}, {Geary},
  {Gould}, {Haas}, {Henze}, {Howard}, {Howell}, {Huber}, {Jenkins}, {Kjeldsen},
  {Kolbl}, {Kolodziejczak}, {Latham}, {Lee}, {Lopez}, {Mullally}, {Orosz},
  {Prsa}, {Quintana}, {Sanchis-Ojeda}, {Sasselov}, {Seader}, {Shporer},
  {Steffen}, {Still}, {Tenenbaum}, {Thompson}, {Torres}, {Twicken}, {Welsh}, \&
  {Winn}}]{borucki2013}
{Borucki}, W.~J., {Agol}, E., {Fressin}, F., {et~al.} 2013, Science, 340, 587

\bibitem[{{Broeg} {et~al.}(2013){Broeg}, {Fortier}, {Ehrenreich}, {Alibert},
  {Baumjohann}, {Benz}, {Deleuil}, {Gillon}, {Ivanov}, {Liseau}, {Meyer},
  {Oloffson}, {Pagano}, {Piotto}, {Pollacco}, {Queloz}, {Ragazzoni}, {Renotte},
  {Steller}, \& {Thomas}}]{broeg2013}
{Broeg}, C., {Fortier}, A., {Ehrenreich}, D., {et~al.} 2013, in European
  Physical Journal Web of Conferences, Vol.~47, European Physical Journal Web
  of Conferences, 3005

\bibitem[{{Cabrera}(2010)}]{cabrera2008}
{Cabrera}, J. 2010, in EAS Publications Series, Vol.~42, EAS Publications
  Series, ed. K.~{Go{\.z}dziewski}, A.~{Niedzielski}, \& J.~{Schneider},
  109--116

\bibitem[{{Cabrera} {et~al.}(2012){Cabrera}, {Csizmadia}, {Erikson}, {Rauer},
  \& {Kirste}}]{cabrera2012}
{Cabrera}, J., {Csizmadia}, S., {Erikson}, A., {Rauer}, H., \& {Kirste}, S.
  2012, \aap, 548, A44

\bibitem[{{Chambers}(1999)}]{chambers1999}
{Chambers}, J.~E. 1999, \mnras, 304, 793

\bibitem[{{Chambers} {et~al.}(1996){Chambers}, {Wetherill}, \&
  {Boss}}]{chambers1996}
{Chambers}, J.~E., {Wetherill}, G.~W., \& {Boss}, A.~P. 1996, \icarus, 119, 261

\bibitem[{{Chaplin} {et~al.}(2013){Chaplin}, {Sanchis-Ojeda}, {Campante},
  {Handberg}, {Stello}, {Winn}, {Basu}, {Christensen-Dalsgaard}, {Davies},
  {Metcalfe}, {Buchhave}, {Fischer}, {Bedding}, {Cochran}, {Elsworth},
  {Gilliland}, {Hekker}, {Huber}, {Isaacson}, {Karoff}, {Kawaler}, {Kjeldsen},
  {Latham}, {Lund}, {Lundkvist}, {Marcy}, {Miglio}, {Barclay}, \&
  {Lissauer}}]{chaplin2013}
{Chaplin}, W.~J., {Sanchis-Ojeda}, R., {Campante}, T.~L., {et~al.} 2013, \apj,
  766, 101

\bibitem[{{Chatterjee} {et~al.}(2008){Chatterjee}, {Ford}, {Matsumura}, \&
  {Rasio}}]{chatterjee2008}
{Chatterjee}, S., {Ford}, E.~B., {Matsumura}, S., \& {Rasio}, F.~A. 2008, \apj,
  686, 580

\bibitem[{{Chatterjee} \& {Tan}(2013)}]{chatterjee2013}
{Chatterjee}, S., \& {Tan}, J.~C. 2013, ArXiv e-prints, arXiv:1306.0576

\bibitem[{{Chiang} \& {Laughlin}(2013)}]{chiang2013}
{Chiang}, E., \& {Laughlin}, G. 2013, \mnras, 431, 3444

\bibitem[{{Cowen}(2013)}]{cowen2013}
{Cowen}, R. 2013, \nat, 497, 417

\bibitem[{{Csizmadia} {et~al.}(2013){Csizmadia}, {Pasternacki}, {Dreyer},
  {Cabrera}, {Erikson}, \& {Rauer}}]{csizmadia2013}
{Csizmadia}, S., {Pasternacki}, T., {Dreyer}, C., {et~al.} 2013, \aap, 549, A9

\bibitem[{{Csizmadia} {et~al.}(2011){Csizmadia}, {Moutou}, {Deleuil},
  {Cabrera}, {Fridlund}, {Gandolfi}, {Aigrain}, {Alonso}, {Almenara},
  {Auvergne}, {Baglin}, {Barge}, {Bonomo}, {Bord{\'e}}, {Bouchy}, {Bruntt},
  {Carone}, {Carpano}, {Cavarroc}, {Cochran}, {Deeg}, {D{\'{\i}}az}, {Dvorak},
  {Endl}, {Erikson}, {Ferraz-Mello}, {Fruth}, {Gazzano}, {Gillon}, {Guenther},
  {Guillot}, {Hatzes}, {Havel}, {H{\'e}brard}, {Jehin}, {Jorda}, {L{\'e}ger},
  {Llebaria}, {Lammer}, {Lovis}, {MacQueen}, {Mazeh}, {Ollivier},
  {P{\"a}tzold}, {Queloz}, {Rauer}, {Rouan}, {Santerne}, {Schneider},
  {Tingley}, {Titz-Weider}, \& {Wuchterl}}]{csizmadia2011}
{Csizmadia}, S., {Moutou}, C., {Deleuil}, M., {et~al.} 2011, \aap, 531, A41

\bibitem[{{Cutri} {et~al.}(2003){Cutri}, {Skrutskie}, {van Dyk}, {Beichman},
  {Carpenter}, {Chester}, {Cambresy}, {Evans}, {Fowler}, {Gizis}, {Howard},
  {Huchra}, {Jarrett}, {Kopan}, {Kirkpatrick}, {Light}, {Marsh}, {McCallon},
  {Schneider}, {Stiening}, {Sykes}, {Weinberg}, {Wheaton}, {Wheelock}, \&
  {Zacarias}}]{cutri2003}
{Cutri}, R.~M., {Skrutskie}, M.~F., {van Dyk}, S., {et~al.} 2003, {2MASS All
  Sky Catalog of point sources.}

\bibitem[{{Deeg} {et~al.}(2010){Deeg}, {Moutou}, {Erikson}, {Csizmadia},
  {Tingley}, {Barge}, {Bruntt}, {Havel}, {Aigrain}, {Almenara}, {Alonso},
  {Auvergne}, {Baglin}, {Barbieri}, {Benz}, {Bonomo}, {Bord{\'e}}, {Bouchy},
  {Cabrera}, {Carone}, {Carpano}, {Ciardi}, {Deleuil}, {Dvorak},
  {Ferraz-Mello}, {Fridlund}, {Gandolfi}, {Gazzano}, {Gillon}, {Gondoin},
  {Guenther}, {Guillot}, {Hartog}, {Hatzes}, {Hidas}, {H{\'e}brard}, {Jorda},
  {Kabath}, {Lammer}, {L{\'e}ger}, {Lister}, {Llebaria}, {Lovis}, {Mayor},
  {Mazeh}, {Ollivier}, {P{\"a}tzold}, {Pepe}, {Pont}, {Queloz}, {Rabus},
  {Rauer}, {Rouan}, {Samuel}, {Schneider}, {Shporer}, {Stecklum}, {Street},
  {Udry}, {Weingrill}, \& {Wuchterl}}]{deeg2010}
{Deeg}, H.~J., {Moutou}, C., {Erikson}, A., {et~al.} 2010, \nat, 464, 384

\bibitem[{{Domingos} {et~al.}(2006){Domingos}, {Winter}, \&
  {Yokoyama}}]{domingos2006}
{Domingos}, R.~C., {Winter}, O.~C., \& {Yokoyama}, T. 2006, \mnras, 373, 1227

\bibitem[{{Dvorak} {et~al.}(1998){Dvorak}, {Contopoulos}, {Efthymiopoulos}, \&
  {Voglis}}]{dvorak1998}
{Dvorak}, R., {Contopoulos}, G., {Efthymiopoulos}, C., \& {Voglis}, N. 1998,
  \planss, 46, 1567

\bibitem[{{Fabrycky} {et~al.}(2012){Fabrycky}, {Ford}, {Steffen}, {Rowe},
  {Carter}, {Moorhead}, {Batalha}, {Borucki}, {Bryson}, {Buchhave},
  {Christiansen}, {Ciardi}, {Cochran}, {Endl}, {Fanelli}, {Fischer}, {Fressin},
  {Geary}, {Haas}, {Hall}, {Holman}, {Jenkins}, {Koch}, {Latham}, {Li},
  {Lissauer}, {Lucas}, {Marcy}, {Mazeh}, {McCauliff}, {Quinn}, {Ragozzine},
  {Sasselov}, \& {Shporer}}]{fabrycky2012a}
{Fabrycky}, D.~C., {Ford}, E.~B., {Steffen}, J.~H., {et~al.} 2012, \apj, 750,
  114

\bibitem[{{Fang} \& {Margot}(2012)}]{fang2012}
{Fang}, J., \& {Margot}, J.-L. 2012, \apj, 761, 92

\bibitem[{{Feroz} {et~al.}(2011){Feroz}, {Balan}, \& {Hobson}}]{feroz2011}
{Feroz}, F., {Balan}, S.~T., \& {Hobson}, M.~P. 2011, \mnras, 415, 3462

\bibitem[{{Ferraz-Mello}(1979)}]{ferrazmello1979}
{Ferraz-Mello}, S. 1979, {Dynamics of the Galilean satellites - an introductory
  treatise}

\bibitem[{{Ford} \& {Gaudi}(2006)}]{ford2006a}
{Ford}, E.~B., \& {Gaudi}, B.~S. 2006, \apjl, 652, L137

\bibitem[{{Ford} {et~al.}(2011){Ford}, {Rowe}, {Fabrycky}, {Carter}, {Holman},
  {Lissauer}, {Ragozzine}, {Steffen}, {Batalha}, {Borucki}, {Bryson},
  {Caldwell}, {Dunham}, {Gautier}, {Jenkins}, {Koch}, {Li}, {Lucas}, {Marcy},
  {McCauliff}, {Mullally}, {Quintana}, {Still}, {Tenenbaum}, {Thompson}, \&
  {Twicken}}]{ford2011}
{Ford}, E.~B., {Rowe}, J.~F., {Fabrycky}, D.~C., {et~al.} 2011, \apjs, 197, 2

\bibitem[{{Ford} {et~al.}(2012{\natexlab{a}}){Ford}, {Fabrycky}, {Steffen},
  {Carter}, {Fressin}, {Holman}, {Lissauer}, {Moorhead}, {Morehead},
  {Ragozzine}, {Rowe}, {Welsh}, {Allen}, {Batalha}, {Borucki}, {Bryson},
  {Buchhave}, {Burke}, {Caldwell}, {Charbonneau}, {Clarke}, {Cochran},
  {D{\'e}sert}, {Endl}, {Everett}, {Fischer}, {Gautier}, {Gilliland},
  {Jenkins}, {Haas}, {Horch}, {Howell}, {Ibrahim}, {Isaacson}, {Koch},
  {Latham}, {Li}, {Lucas}, {MacQueen}, {Marcy}, {McCauliff}, {Mullally},
  {Quinn}, {Quintana}, {Shporer}, {Still}, {Tenenbaum}, {Thompson}, {Torres},
  {Twicken}, {Wohler}, \& {Kepler Science Team}}]{ford2012a}
{Ford}, E.~B., {Fabrycky}, D.~C., {Steffen}, J.~H., {et~al.}
  2012{\natexlab{a}}, \apj, 750, 113

\bibitem[{{Ford} {et~al.}(2012{\natexlab{b}}){Ford}, {Ragozzine}, {Rowe},
  {Steffen}, {Barclay}, {Batalha}, {Borucki}, {Bryson}, {Caldwell}, {Fabrycky},
  {Gautier}, {Holman}, {Ibrahim}, {Kjeldsen}, {Kinemuchi}, {Koch}, {Lissauer},
  {Still}, {Tenenbaum}, {Uddin}, \& {Welsh}}]{ford2012b}
{Ford}, E.~B., {Ragozzine}, D., {Rowe}, J.~F., {et~al.} 2012{\natexlab{b}},
  \apj, 756, 185

\bibitem[{{Fressin} {et~al.}(2012){Fressin}, {Torres}, {Rowe}, {Charbonneau},
  {Rogers}, {Ballard}, {Batalha}, {Borucki}, {Bryson}, {Buchhave}, {Ciardi},
  {D{\'e}sert}, {Dressing}, {Fabrycky}, {Ford}, {Gautier}, {Henze}, {Holman},
  {Howard}, {Howell}, {Jenkins}, {Koch}, {Latham}, {Lissauer}, {Marcy},
  {Quinn}, {Ragozzine}, {Sasselov}, {Seager}, {Barclay}, {Mullally}, {Seader},
  {Still}, {Twicken}, {Thompson}, \& {Uddin}}]{fressin2012a}
{Fressin}, F., {Torres}, G., {Rowe}, J.~F., {et~al.} 2012, \nat, 482, 195

\bibitem[{{Gandolfi} {et~al.}(2008){Gandolfi}, {Alcal{\'a}}, {Leccia},
  {Frasca}, {Spezzi}, {Covino}, {Testi}, {Marilli}, \&
  {Kainulainen}}]{gandolfi2008}
{Gandolfi}, D., {Alcal{\'a}}, J.~M., {Leccia}, S., {et~al.} 2008, \apj, 687,
  1303

\bibitem[{{Gautier} {et~al.}(2012){Gautier}, {Charbonneau}, {Rowe}, {Marcy},
  {Isaacson}, {Torres}, {Fressin}, {Rogers}, {D{\'e}sert}, {Buchhave},
  {Latham}, {Quinn}, {Ciardi}, {Fabrycky}, {Ford}, {Gilliland}, {Walkowicz},
  {Bryson}, {Cochran}, {Endl}, {Fischer}, {Howell}, {Horch}, {Barclay},
  {Batalha}, {Borucki}, {Christiansen}, {Geary}, {Henze}, {Holman}, {Ibrahim},
  {Jenkins}, {Kinemuchi}, {Koch}, {Lissauer}, {Sanderfer}, {Sasselov},
  {Seager}, {Silverio}, {Smith}, {Still}, {Stumpe}, {Tenenbaum}, \& {Van
  Cleve}}]{gautier2012}
{Gautier}, III, T.~N., {Charbonneau}, D., {Rowe}, J.~F., {et~al.} 2012, \apj,
  749, 15

\bibitem[{{Geem} {et~al.}(2001){Geem}, {Kim}, \& {Loganathan}}]{geem2001}
{Geem}, Z.~G., {Kim}, J.~H., \& {Loganathan}, G.~V. 2001, Simulation, 76, 60,
  http://sim.sagepub.com/cgi/content/abstract/76/2/60

\bibitem[{{Gim{\'e}nez}(2006)}]{gimenez2006}
{Gim{\'e}nez}, A. 2006, \aap, 450, 1231

\bibitem[{{Goldreich} \& {Tremaine}(1980)}]{goldreich1980}
{Goldreich}, P., \& {Tremaine}, S. 1980, \apj, 241, 425

\bibitem[{{Gong} {et~al.}(2013){Gong}, {Zhou}, {Xie}, \& {Wu}}]{gong2013}
{Gong}, Y.-X., {Zhou}, J.-L., {Xie}, J.-W., \& {Wu}, X.-M. 2013, \apjl, 769,
  L14

\bibitem[{{Gray}(2005)}]{gray2005}
{Gray}, D.~F. 2005, {The Observation and Analysis of Stellar Photospheres}

\bibitem[{{Hanslmeier} \& {Dvorak}(1984)}]{hanslmeier1984}
{Hanslmeier}, A., \& {Dvorak}, R. 1984, \aap, 132, 203

\bibitem[{{Hatzes}(2013)}]{hatzes2013b}
{Hatzes}, A.~P. 2013, Astronomische Nachrichten, 334, 616

\bibitem[{{Hauschildt} {et~al.}(1999){Hauschildt}, {Allard}, \&
  {Baron}}]{hauschildt1999a}
{Hauschildt}, P.~H., {Allard}, F., \& {Baron}, E. 1999, \apj, 512, 377

\bibitem[{{Holman} \& {Murray}(2005)}]{holman2005}
{Holman}, M.~J., \& {Murray}, N.~W. 2005, Science, 307, 1288

\bibitem[{{Huang} {et~al.}(2013){Huang}, {Bakos}, \& {Hartman}}]{huang2013}
{Huang}, X., {Bakos}, G.~{\'A}., \& {Hartman}, J.~D. 2013, \mnras, 429, 2001

\bibitem[{{Kipping}(2009{\natexlab{a}})}]{kipping2009a}
{Kipping}, D.~M. 2009{\natexlab{a}}, \mnras, 392, 181

\bibitem[{{Kipping}(2009{\natexlab{b}})}]{kipping2009b}
---. 2009{\natexlab{b}}, \mnras, 396, 1797

\bibitem[{{Kipping}(2011)}]{kipping2011b}
---. 2011, \mnras, 416, 689

\bibitem[{{Kipping} {et~al.}(2012){Kipping}, {Bakos}, {Buchhave},
  {Nesvorn{\'y}}, \& {Schmitt}}]{kipping2012c}
{Kipping}, D.~M., {Bakos}, G.~{\'A}., {Buchhave}, L., {Nesvorn{\'y}}, D., \&
  {Schmitt}, A. 2012, \apj, 750, 115

\bibitem[{{Kipping} {et~al.}(2013{\natexlab{a}}){Kipping}, {Forgan}, {Hartman},
  {Nesvorny}, {Bakos}, {Schmitt}, \& {Buchhave}}]{kipping2013b}
{Kipping}, D.~M., {Forgan}, D., {Hartman}, J., {et~al.} 2013{\natexlab{a}},
  ArXiv e-prints, arXiv:1306.1530

\bibitem[{{Kipping} {et~al.}(2013{\natexlab{b}}){Kipping}, {Hartman},
  {Buchhave}, {Schmitt}, {Bakos}, \& {Nesvorn{\'y}}}]{kipping2013a}
{Kipping}, D.~M., {Hartman}, J., {Buchhave}, L.~A., {et~al.}
  2013{\natexlab{b}}, \apj, 770, 101

\bibitem[{{Laskar}(1994)}]{laskar1994}
{Laskar}, J. 1994, \aap, 287, L9

\bibitem[{{Laskar}(2008)}]{laskar2008}
---. 2008, \icarus, 196, 1

\bibitem[{{L{\'e}ger} {et~al.}(2009){L{\'e}ger}, {Rouan}, {Schneider}, {Barge},
  {Fridlund}, {Samuel}, {Ollivier}, {Guenther}, {Deleuil}, {Deeg}, {Auvergne},
  {Alonso}, {Aigrain}, {Alapini}, {Almenara}, {Baglin}, {Barbieri}, {Bruntt},
  {Bord{\'e}}, {Bouchy}, {Cabrera}, {Catala}, {Carone}, {Carpano}, {Csizmadia},
  {Dvorak}, {Erikson}, {Ferraz-Mello}, {Foing}, {Fressin}, {Gandolfi},
  {Gillon}, {Gondoin}, {Grasset}, {Guillot}, {Hatzes}, {H{\'e}brard}, {Jorda},
  {Lammer}, {Llebaria}, {Loeillet}, {Mayor}, {Mazeh}, {Moutou}, {P{\"a}tzold},
  {Pont}, {Queloz}, {Rauer}, {Renner}, {Samadi}, {Shporer}, {Sotin}, {Tingley},
  {Wuchterl}, {Adda}, {Agogu}, {Appourchaux}, {Ballans}, {Baron}, {Beaufort},
  {Bellenger}, {Berlin}, {Bernardi}, {Blouin}, {Baudin}, {Bodin}, {Boisnard},
  {Boit}, {Bonneau}, {Borzeix}, {Briet}, {Buey}, {Butler}, {Cailleau},
  {Cautain}, {Chabaud}, {Chaintreuil}, {Chiavassa}, {Costes}, {Cuna Parrho},
  {de Oliveira Fialho}, {Decaudin}, {Defise}, {Djalal}, {Epstein}, {Exil},
  {Faur{\'e}}, {Fenouillet}, {Gaboriaud}, {Gallic}, {Gamet}, {Gavalda},
  {Grolleau}, {Gruneisen}, {Gueguen}, {Guis}, {Guivarc'h}, {Guterman},
  {Hallouard}, {Hasiba}, {Heuripeau}, {Huntzinger}, {Hustaix}, {Imad},
  {Imbert}, {Johlander}, {Jouret}, {Journoud}, {Karioty}, {Kerjean},
  {Lafaille}, {Lafond}, {Lam-Trong}, {Landiech}, {Lapeyrere}, {Larqu{\'e}},
  {Laudet}, {Lautier}, {Lecann}, {Lefevre}, {Leruyet}, {Levacher}, {Magnan},
  {Mazy}, {Mertens}, {Mesnager}, {Meunier}, {Michel}, {Monjoin}, {Naudet},
  {Nguyen-Kim}, {Orcesi}, {Ottacher}, {Perez}, {Peter}, {Plasson}, {Plesseria},
  {Pontet}, {Pradines}, {Quentin}, {Reynaud}, {Rolland}, {Rollenhagen},
  {Romagnan}, {Russ}, {Schmidt}, {Schwartz}, {Sebbag}, {Sedes}, {Smit},
  {Steller}, {Sunter}, {Surace}, {Tello}, {Tiph{\`e}ne}, {Toulouse}, {Ulmer},
  {Vandermarcq}, {Vergnault}, {Vuillemin}, \& {Zanatta}}]{leger2009}
{L{\'e}ger}, A., {Rouan}, D., {Schneider}, J., {et~al.} 2009, \aap, 506, 287

\bibitem[{{Lehmann} {et~al.}(2011){Lehmann}, {Tkachenko}, {Semaan},
  {Guti{\'e}rrez-Soto}, {Smalley}, {Briquet}, {Shulyak}, {Tsymbal}, \& {De
  Cat}}]{lehman2011}
{Lehmann}, H., {Tkachenko}, A., {Semaan}, T., {et~al.} 2011, \aap, 526, A124

\bibitem[{{Lin} {et~al.}(1996){Lin}, {Bodenheimer}, \& {Richardson}}]{lin1996}
{Lin}, D.~N.~C., {Bodenheimer}, P., \& {Richardson}, D.~C. 1996, \nat, 380, 606

\bibitem[{{Lin} \& {Ida}(1997)}]{lin1997}
{Lin}, D.~N.~C., \& {Ida}, S. 1997, \apj, 477, 781

\bibitem[{{Lissauer} {et~al.}(2011){Lissauer}, {Fabrycky}, {Ford}, {Borucki},
  {Fressin}, {Marcy}, {Orosz}, {Rowe}, {Torres}, {Welsh}, {Batalha}, {Bryson},
  {Buchhave}, {Caldwell}, {Carter}, {Charbonneau}, {Christiansen}, {Cochran},
  {Desert}, {Dunham}, {Fanelli}, {Fortney}, {Gautier}, {Geary}, {Gilliland},
  {Haas}, {Hall}, {Holman}, {Koch}, {Latham}, {Lopez}, {McCauliff}, {Miller},
  {Morehead}, {Quintana}, {Ragozzine}, {Sasselov}, {Short}, \&
  {Steffen}}]{lissauer2011a}
{Lissauer}, J.~J., {Fabrycky}, D.~C., {Ford}, E.~B., {et~al.} 2011, \nat, 470,
  53

\bibitem[{{Lissauer} {et~al.}(2012){Lissauer}, {Marcy}, {Rowe}, {Bryson},
  {Adams}, {Buchhave}, {Ciardi}, {Cochran}, {Fabrycky}, {Ford}, {Fressin},
  {Geary}, {Gilliland}, {Holman}, {Howell}, {Jenkins}, {Kinemuchi}, {Koch},
  {Morehead}, {Ragozzine}, {Seader}, {Tanenbaum}, {Torres}, \&
  {Twicken}}]{lissauer2012a}
{Lissauer}, J.~J., {Marcy}, G.~W., {Rowe}, J.~F., {et~al.} 2012, \apj, 750, 112

\bibitem[{{Lithwick} \& {Wu}(2012)}]{lithwick2012b}
{Lithwick}, Y., \& {Wu}, Y. 2012, \apjl, 756, L11

\bibitem[{{Lithwick} {et~al.}(2012){Lithwick}, {Xie}, \& {Wu}}]{lithwick2012a}
{Lithwick}, Y., {Xie}, J., \& {Wu}, Y. 2012, \apj, 761, 122

\bibitem[{{Lovis} {et~al.}(2011){Lovis}, {S{\'e}gransan}, {Mayor}, {Udry},
  {Benz}, {Bertaux}, {Bouchy}, {Correia}, {Laskar}, {Lo Curto}, {Mordasini},
  {Pepe}, {Queloz}, \& {Santos}}]{lovis2011}
{Lovis}, C., {S{\'e}gransan}, D., {Mayor}, M., {et~al.} 2011, \aap, 528, A112

\bibitem[{{Mandel} \& {Agol}(2002)}]{mandel2002}
{Mandel}, K., \& {Agol}, E. 2002, \apjl, 580, L171

\bibitem[{{Mazeh} {et~al.}(2013){Mazeh}, {Nachmani}, {Holczer}, {Fabrycky},
  {Ford}, {Sanchis-Ojeda}, {Sokol}, {Rowe}, {Zucker}, {Agol}, {Carter},
  {Lissauer}, {Quintana}, {Ragozzine}, {Steffen}, \& {Welsh}}]{mazeh2013}
{Mazeh}, T., {Nachmani}, G., {Holczer}, T., {et~al.} 2013, \apjs, 208, 16

\bibitem[{{Murray} {et~al.}(1998){Murray}, {Hansen}, {Holman}, \&
  {Tremaine}}]{murray1998}
{Murray}, N., {Hansen}, B., {Holman}, M., \& {Tremaine}, S. 1998, Science, 279,
  69

\bibitem[{{Namouni}(2010)}]{namouni2010}
{Namouni}, F. 2010, \apjl, 719, L145

\bibitem[{{Nesvorn{\'y}} {et~al.}(2013){Nesvorn{\'y}}, {Kipping}, {Terrell},
  {Hartman}, {Bakos}, \& {Buchhave}}]{nesvorny2013a}
{Nesvorn{\'y}}, D., {Kipping}, D., {Terrell}, D., {et~al.} 2013, \apj, 777, 3

\bibitem[{{Ofir} \& {Dreizler}(2013)}]{ofir2013a}
{Ofir}, A., \& {Dreizler}, S. 2013, \aap, 555, A58

\bibitem[{{Petrovich} {et~al.}(2013){Petrovich}, {Malhotra}, \&
  {Tremaine}}]{petrovich2013}
{Petrovich}, C., {Malhotra}, R., \& {Tremaine}, S. 2013, \apj, 770, 24

\bibitem[{{Pont} {et~al.}(2007){Pont}, {Gilliland}, {Moutou}, {Charbonneau},
  {Bouchy}, {Brown}, {Mayor}, {Queloz}, {Santos}, \& {Udry}}]{pont2007}
{Pont}, F., {Gilliland}, R.~L., {Moutou}, C., {et~al.} 2007, \aap, 476, 1347

\bibitem[{{Rasio} \& {Ford}(1996)}]{rasio1996}
{Rasio}, F.~A., \& {Ford}, E.~B. 1996, Science, 274, 954

\bibitem[{{Rauer} \& {Catala}(2011)}]{rauer2011b}
{Rauer}, H., \& {Catala}, C. 2011, in IAU Symposium, Vol. 276, IAU Symposium,
  ed. A.~{Sozzetti}, M.~G. {Lattanzi}, \& A.~P. {Boss}, 354--358

\bibitem[{{Raymond} {et~al.}(2008){Raymond}, {Barnes}, \&
  {Mandell}}]{raymond2008}
{Raymond}, S.~N., {Barnes}, R., \& {Mandell}, A.~M. 2008, \mnras, 384, 663

\bibitem[{{Ricker} {et~al.}(2010){Ricker}, {Latham}, {Vanderspek}, {Ennico},
  {Bakos}, {Brown}, {Burgasser}, {Charbonneau}, {Clampin}, {Deming}, {Doty},
  {Dunham}, {Elliot}, {Holman}, {Ida}, {Jenkins}, {Jernigan}, {Kawai},
  {Laughlin}, {Lissauer}, {Martel}, {Sasselov}, {Schingler}, {Seager},
  {Torres}, {Udry}, {Villasenor}, {Winn}, \& {Worden}}]{ricker2010}
{Ricker}, G.~R., {Latham}, D.~W., {Vanderspek}, R.~K., {et~al.} 2010, in
  Bulletin of the American Astronomical Society, Vol.~42, American Astronomical
  Society Meeting Abstracts 215, 450.06

\bibitem[{{Rivera} {et~al.}(2010){Rivera}, {Laughlin}, {Butler}, {Vogt},
  {Haghighipour}, \& {Meschiari}}]{rivera2010}
{Rivera}, E.~J., {Laughlin}, G., {Butler}, R.~P., {et~al.} 2010, \apj, 719, 890

\bibitem[{{Sartoretti} \& {Schneider}(1999)}]{sartoretti1999}
{Sartoretti}, P., \& {Schneider}, J. 1999, \aaps, 134, 553

\bibitem[{{Shulyak} {et~al.}(2004){Shulyak}, {Tsymbal}, {Ryabchikova},
  {St{\"u}tz}, \& {Weiss}}]{shulyak2004}
{Shulyak}, D., {Tsymbal}, V., {Ryabchikova}, T., {St{\"u}tz}, C., \& {Weiss},
  W.~W. 2004, \aap, 428, 993

\bibitem[{{Steffen} {et~al.}(2012{\natexlab{a}}){Steffen}, {Fabrycky}, {Ford},
  {Carter}, {D{\'e}sert}, {Fressin}, {Holman}, {Lissauer}, {Moorhead}, {Rowe},
  {Ragozzine}, {Welsh}, {Batalha}, {Borucki}, {Buchhave}, {Bryson}, {Caldwell},
  {Charbonneau}, {Ciardi}, {Cochran}, {Endl}, {Everett}, {Gautier},
  {Gilliland}, {Girouard}, {Jenkins}, {Horch}, {Howell}, {Isaacson}, {Klaus},
  {Koch}, {Latham}, {Li}, {Lucas}, {MacQueen}, {Marcy}, {McCauliff}, {Middour},
  {Morris}, {Mullally}, {Quinn}, {Quintana}, {Shporer}, {Still}, {Tenenbaum},
  {Thompson}, {Twicken}, \& {Van Cleve}}]{steffen2012a}
{Steffen}, J.~H., {Fabrycky}, D.~C., {Ford}, E.~B., {et~al.}
  2012{\natexlab{a}}, \mnras, 421, 2342

\bibitem[{{Steffen} {et~al.}(2012{\natexlab{b}}){Steffen}, {Ford}, {Rowe},
  {Fabrycky}, {Holman}, {Welsh}, {Batalha}, {Borucki}, {Bryson}, {Caldwell},
  {Ciardi}, {Jenkins}, {Kjeldsen}, {Koch}, {Pr{\v s}a}, {Sanderfer}, {Seader},
  \& {Twicken}}]{steffen2012b}
{Steffen}, J.~H., {Ford}, E.~B., {Rowe}, J.~F., {et~al.} 2012{\natexlab{b}},
  \apj, 756, 186

\bibitem[{{Steffen} {et~al.}(2013){Steffen}, {Fabrycky}, {Agol}, {Ford},
  {Morehead}, {Cochran}, {Lissauer}, {Adams}, {Borucki}, {Bryson}, {Caldwell},
  {Dupree}, {Jenkins}, {Robertson}, {Rowe}, {Seader}, {Thompson}, \&
  {Twicken}}]{steffen2013}
{Steffen}, J.~H., {Fabrycky}, D.~C., {Agol}, E., {et~al.} 2013, \mnras, 428,
  1077

\bibitem[{{Szab{\'o}} {et~al.}(2006){Szab{\'o}}, {Szatm{\'a}ry}, {Div{\'e}ki},
  \& {Simon}}]{szabo2006}
{Szab{\'o}}, G.~M., {Szatm{\'a}ry}, K., {Div{\'e}ki}, Z., \& {Simon}, A. 2006,
  \aap, 450, 395

\bibitem[{{Tenenbaum} {et~al.}(2013){Tenenbaum}, {Jenkins}, {Seader}, {Burke},
  {Christiansen}, {Rowe}, {Caldwell}, {Clarke}, {Li}, {Quintana}, {Smith},
  {Thompson}, {Twicken}, {Borucki}, {Batalha}, {Cote}, {Haas}, {Hunter},
  {Sanderfer}, {Girouard}, {Hall}, {Ibrahim}, {Klaus}, {McCauliff}, {Middour},
  {Sabale}, {Uddin}, {Wohler}, {Barclay}, \& {Still}}]{tenenbaum2013}
{Tenenbaum}, P., {Jenkins}, J.~M., {Seader}, S., {et~al.} 2013, \apjs, 206, 5

\bibitem[{{Tremaine} \& {Dong}(2012)}]{tremaine2012}
{Tremaine}, S., \& {Dong}, S. 2012, \aj, 143, 94

\bibitem[{{Tsymbal}(1996)}]{tsymbal1996}
{Tsymbal}, V. 1996, in Astronomical Society of the Pacific Conference Series,
  Vol. 108, M.A.S.S., Model Atmospheres and Spectrum Synthesis, ed. S.~J.
  {Adelman}, F.~{Kupka}, \& W.~W. {Weiss}, 198

\bibitem[{{Tuomi}(2012)}]{tuomi2012}
{Tuomi}, M. 2012, \aap, 543, A52

\bibitem[{{Tuomi} {et~al.}(2013){Tuomi}, {Anglada-Escud{\'e}}, {Gerlach},
  {Jones}, {Reiners}, {Rivera}, {Vogt}, \& {Butler}}]{tuomi2013}
{Tuomi}, M., {Anglada-Escud{\'e}}, G., {Gerlach}, E., {et~al.} 2013, \aap, 549,
  A48

\bibitem[{{Wang} {et~al.}(2012){Wang}, {Ji}, \& {Zhou}}]{wang2012b}
{Wang}, S., {Ji}, J., \& {Zhou}, J.-L. 2012, \apj, 753, 170

\bibitem[{{Ward}(1997)}]{ward1997}
{Ward}, W.~R. 1997, Icarus, 126, 261

\bibitem[{{Wright} {et~al.}(2010){Wright}, {Eisenhardt}, {Mainzer}, {Ressler},
  {Cutri}, {Jarrett}, {Kirkpatrick}, {Padgett}, {McMillan}, {Skrutskie},
  {Stanford}, {Cohen}, {Walker}, {Mather}, {Leisawitz}, {Gautier}, {McLean},
  {Benford}, {Lonsdale}, {Blain}, {Mendez}, {Irace}, {Duval}, {Liu}, {Royer},
  {Heinrichsen}, {Howard}, {Shannon}, {Kendall}, {Walsh}, {Larsen}, {Cardon},
  {Schick}, {Schwalm}, {Abid}, {Fabinsky}, {Naes}, \& {Tsai}}]{wright2010wise}
{Wright}, E.~L., {Eisenhardt}, P.~R.~M., {Mainzer}, A.~K., {et~al.} 2010, \aj,
  140, 1868

\bibitem[{{Wu} \& {Lithwick}(2013)}]{wu2013}
{Wu}, Y., \& {Lithwick}, Y. 2013, \apj, 772, 74

\bibitem[{{Xie}(2013)}]{xie2013}
{Xie}, J.-W. 2013, \apjs, 208, 22

\end{thebibliography}

\end{document}